\documentclass[sigconf]{acmart}

\AtBeginDocument{%
  \providecommand\BibTeX{{%
    \normalfont B\kern-0.5em{\scshape i\kern-0.25em b}\kern-0.8em\TeX}}}

\setcopyright{acmcopyright}

\copyrightyear{2023}
\acmYear{2023}
\setcopyright{acmlicensed}\acmConference[MobiSys '23]{The 21st Annual International Conference on Mobile Systems, Applications and Services}{June 18--22, 2023}{Helsinki, Finland}
\acmBooktitle{The 21st Annual International Conference on Mobile Systems, Applications and Services (MobiSys '23), June 18--22, 2023, Helsinki, Finland}
\acmPrice{15.00}
\acmDOI{10.1145/3581791.3596859}
\acmISBN{979-8-4007-0110-8/23/06}

\usepackage{lipsum}
  \usepackage{amsmath,amsfonts}
 \usepackage{graphicx}
 \usepackage{textcomp}
 \usepackage{xspace}
 \usepackage{xcolor}
 \usepackage{lipsum} 
 \usepackage[utf8]{inputenc}
 \usepackage{siunitx}
 \usepackage{algorithm, algpseudocode}
 \usepackage{soul}

\usepackage{mathtools}
\usepackage{xcolor}
\definecolor{mycolor}{RGB}{0,140,0}
\hypersetup{
    colorlinks=true,
    linkcolor=black,
    citecolor=black,    
    urlcolor=mycolor,
}
\usepackage{comment}
\usepackage{url}
\usepackage{subcaption}
\usepackage{caption}
\usepackage{threeparttable}
\usepackage{balance}
\usepackage{textcomp, libertine}

\usepackage[font=small,skip=2pt]{caption}

\newcommand{\parlabel}[1]{\vspace{0.2em}\noindent\textbf{#1}}

\DeclareMathOperator*{\argmin}{\arg\!\min}

\newcommand{\sys}{SoundSieve\xspace}
\newcommand{\SYS}{SOUNDSIEVE\xspace}

\begin{document}

\title{\sys: Seconds-Long Audio Event Recognition on Intermittently-Powered Systems}
\author{Mahathir Monjur, Yubo Luo, Zhenyu Wang, Shahriar Nirjon}
\affiliation{%
  \institution{Department of Computer Science\\University of North Carolina at Chapel Hill}
   \city{Chapel Hill}
   \state{NC}
   \country{USA}
}
\email{{mahathir, yubo, zywang, nirjon}@cs.unc.edu}

\begin{abstract} 
A fundamental problem of every intermittently-powered sensing system is that signals acquired by these systems over a longer period in time are also \emph{intermittent}. As a consequence, these systems fail to capture parts of a longer-duration event that spans over multiple charge-discharge cycles of the capacitor that stores the harvested energy. From an application's perspective, this is viewed as sporadic bursts of missing values in the input data -- which may not be recoverable using statistical interpolation or imputation methods. In this paper, we study this problem in the light of an intermittent audio classification system and design an end-to-end system -- \emph{\sys} -- that is capable of accurately classifying audio events that span multiple on-off cycles of the intermittent system. \sys employs an offline \emph{audio analyzer} that learns to identify and predict important segments of an audio clip that must be sampled to ensure accurate classification of the audio. At runtime, \sys employs a lightweight, energy- and content-aware \emph{audio sampler} that decides when the system should wake up to capture the next chunk of audio; and a lightweight, intermittence-aware \emph{audio classifier} that performs imputation and on-device inference. Through extensive evaluations using popular audio datasets as well as real systems, we demonstrate that \sys yields 5\%--30\% more accurate inference results than the state-of-the-art.                                  

\end{abstract}



\maketitle

\section{Introduction}
\label{sec:intro}

As batteryless computing systems continue to mature, we see a gradual shift in research from developing tools and platforms for energy harvesting systems~\cite{ flicker, colin2016energy, norenberg2021stonehenge, sigrist2020environment, hadas2018virtual}, to devising new programming paradigms and runtime systems~\cite{islam2020scheduling, colin2016chain, colin2018termination, hester2017timely}, to more recently, implementing machine learning techniques and applications~\cite{bashima2019ondevice, luo2023efficient, islam2019zygarde, lee2019neuro, wang2021lightweight} tailored to intermittently-powered systems that run on harvested energy from solar, kinetic, thermal, or RF sources. A wide variety of on-device inference-capable intermittent systems have been proposed in the recent literature~\cite{continuous, islam2023amalgamated, mishra2021origin, gobieski2019intelligence, lee2019intermittent, lee2019neuro, campbell2014energy, spoton, smarton, wu2022towards, 9111002, tang2022enabling, dong2021semi} that perform on-device inference of audio, image, environmental parameters, building activity, and human activity recognition tasks.   


\begin{figure}[!htb]
    \centering
    \vspace{-5pt}
    \includegraphics[height=1.5in, width=\linewidth]{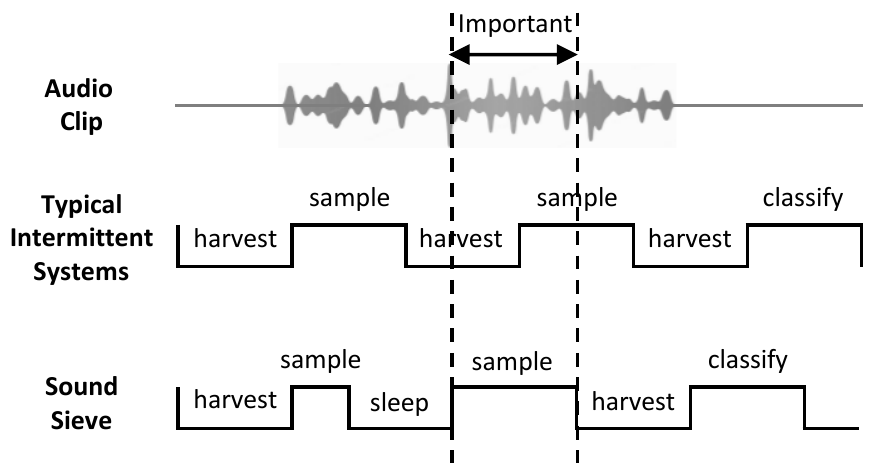}
    \caption{A typical intermittent system, for lack of content awareness, misses important portions of the audio. \sys samples and analyzes an earlier segment, decides (in real-time) to conserve energy by sleeping (and also by harvesting while in sleep mode) and use that energy later to capture the important segment. Note that there can be multiple non-consecutive important segments in an audio. We only show one for illustration.}
    \label{fig:page1}
    \vspace{-10pt}
\end{figure}

Unfortunately, existing intermittent systems have only been demonstrated successful in classifying short-lived events that last for about a few hundred milliseconds~\cite{continuous}. This limit is fundamental to the design of intermittent systems that require a fixed-sized capacitor to accumulate enough harvested energy for the system to wake up and remain active. The system in its active mode consumes the harvested energy to perform application-specific tasks, but soon runs out of energy and goes back to sleep. By choosing a larger (or smaller) capacitor, the lengths of the active and sleep phases can be increased (or decreased), but generally, there will always be one or more sleep phases when the system is unable to acquire any sensor data. This is why state-of-the-art intermittent systems implicitly assume that the event of interest lasts no more than one discharging cycle of the capacitor so that it can be captured fully in one active phase of the system. Events that span across multiple charge-discharge cycles of the capacitor are not fully captured by existing intermittent systems. Hence, specialized techniques are required to ensure accurate classification of longer-duration events on intermittently-powered systems.



In this paper, we take up the challenge of audio event classification on intermittent systems where events may last longer than one charging cycle (and possibly more). To achieve this, we exploit two empirically-learned properties of audio signals: first, \emph{relevance} -- i.e., not every segment of an audio clip contributes equally to its classification; and second, \emph{imputability} -- i.e., many segments within an audio clip can be imputed or estimated to a sufficient detail from other nearby segments. 

These observations lead us to the design of \emph{\sys} --- a software-only, content- and energy-aware solution to seconds-long audio event classification on intermittent systems. The key idea behind \sys is its ability to predict and control when an intermittent system should wake up from sleep -- in order to capture and process only the \emph{relevant, non-imputable} segments of the input audio. Unlike existing intermittent systems that continuously sample and process audio until the capacitor is fully drained of harvested energy, \sys saves energy by switching to sleep mode when the incoming audio does not carry information that is new or relevant to classification. This strategy increases \sys's discharge time since the capacitor now can hold charge and use that energy later. Because of this, \sys is able to sense and process input signals during intervals when existing intermittent systems cannot. This contrast is illustrated in Figure~\ref{fig:page1}.             

Although the idea of \sys is simple and intuitive, developing such a solution is challenging due to severe CPU, memory, and energy constraints of an intermittent system. \sys cannot afford computationally expensive sampling techniques that are suitable for battery-powered, advanced mobile systems such as robotic explorers~\cite{kemna2016adaptive, fung2019coordinating, wu2020enabling}. \sys must sample a small segment of the input audio, analyze its content, and  decide whether to continue sampling or to go to sleep (and if so, for how long). All of these must happen in real-time, and ideally, without any additional energy overhead. \sys achieves this through a combination of offline and online algorithms. 

\sys employs an offline \emph{audio analyzer} that learns to identify and predict important segments of an audio clip that must be sampled to ensure accurate classification of the audio clip. At runtime, \sys employs a lightweight, energy- and content-aware \emph{audio sampler} that decides when the system should wake up to capture the next audio segment. The sampler's decision algorithm runs in parallel with the signal acquisition process to ensure that there is no gap in signal acquisition. The decision algorithm purposefully applies a convolution filter on the sampled audio to make accurate sampling decisions based on high-level acoustic features. Since this filter is also the first layer of the classifier (by design), there is practically no extra energy overhead of the audio sampler. Once an audio event ends, a lightweight, intermittence-aware \emph{audio classifier} performs imputation and classification.

Several salient features make \sys unique of its kind. First, \sys is the first intermittent system that makes content-aware sampling and processing decisions to enable long duration audio event recognition. Second, \sys is a single-node, software-only solution that neither requires multiple coordinated harvesters, nor any modification to the harvester, nor any additional hardware components beyond a basic intermittent computing system. Third, \sys is agnostic to the harvester and intermittence pattern. Even battery-powered systems can adopt \sys's intelligent sensing mechanism to extend their battery-life. Fourth, \sys's signal processing framework is generalizable beyond audio. It can inspire future intermittent systems that detect complex patterns in time-series signals~\cite{zhang2022let,zhang2023spy} such as video~\cite{liu2021video} and motion sensors~\cite{liu2021neuropose,zhou2022learning,liu2022leveraging}.




In order to evaluate \sys, we conduct dataset-driven experiments as well as real-world deployments with these systems. In the dataset-driven experiments, we compare the performance of \sys against three baseline solutions, including state-of-the-art intermittent system~\cite{continuous}, over four datasets that contain over 140,000 audio clips from over 125 categories of sounds having the duration of up to 4 seconds. We observe that \sys outperforms all baselines by a significant margin --- \sys successfully captures 5\%--30\% more (relevant) audio segments, and as a result, achieves 5\%--25\% higher inference accuracy than the baseline solutions. For the real-world deployments, we develop a 16-bit TI MSP430FR5994-based intermittent system that senses audio in real-time and performs on-device inference while being powered by solar or RF energy. We implement two applications that involve recognizing voice commands and household activities, respectively. We demonstrate that \sys  captures and accurately classifies 20\% more events than the baseline intermittent systems which do not perform content-aware sampling. 




    


\section{Empirical Study}


In order to understand the effect of missing values caused by intermittence, we conduct an empirical study on three popular audio datasets: Urban8K~\cite{urban}, ESC-50~\cite{piczak2015esc}, and GSC~\cite{warden2018speech}. These datasets contain over 115,000 audio clips of over 95 different categories.

\subsection{Dealing With Missing Values}

\begin{figure}[!htb]
    \centering
    \vspace{-5pt}
    \includegraphics[width=0.75\linewidth]{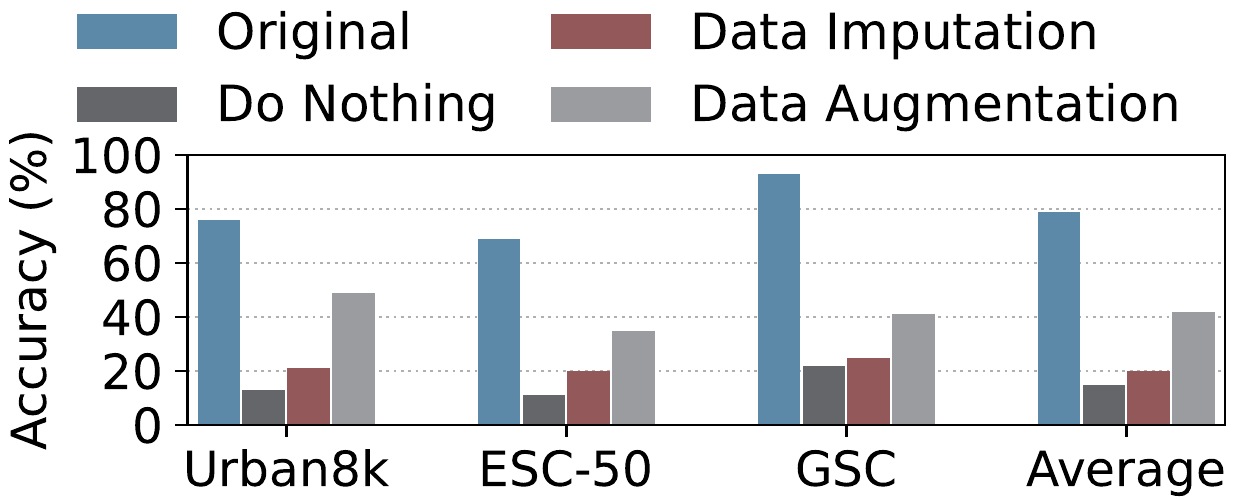}
    \caption{Imputation and augmentation increase accuracy, but cannot maximize the gain in accuracy by themselves. The percentage of missing values in the experiment is $75\%$.}
    \label{fig:iasa}
\end{figure}

A classifier that does not explicitly handle missing values performs extremely poorly when it encounters such inputs. This is evident in Figure~\ref{fig:iasa} -- confirming similar prior studies~\cite{perforation} -- where we see 20\%--50\% drop in inference accuracy (referred to as \emph{do nothing}) when $75\%$ of the data are missing. 

To address this issue, there are two generic classes of solutions in the literature: (a) estimating (imputing) the missing values (before classification) using statistical or neural methods, and (b) training the classifier on an augmented dataset where missing values are synthetically introduced in the training examples. Although implementing any or both of these techniques improve the classification accuracy -- referred to as \emph{imputation} and \emph{augmentation}, respectively --  there is still a large gap of 5\%--40\% in accuracy that cannot be regained. Furthermore, the classifier used in this experiment is a state-of-the-art CNN having 1008 filters, which is not feasible for intermittent systems. In conclusion, \textbf{data imputation and/or data augmentation cannot completely solve the intermittent audio classification problem. We need to complement them with audio content-aware intelligent sampling at the front to maximize their performance.}


\subsection{Feasibility of Content-Aware Sampling}

Although an intermittent system does not have complete control over the harvesting (charging) cycles, which depend on the energy source, once charged, the system can decide when to use that harvested energy. This creates an opportunity for the system to selectively sample segments of the audio that are more informative and relevant to audio classification. 

We conduct another study to understand what percentage of an audio clip must be sampled so that we have enough information to classify a clip correctly. To conduct this study, we (logically) divide each clip into non-overlapping 100ms \emph{segments} and  search for the smallest set of \emph{K} most informative segments that must be sampled to correctly classify the clip. We keep the \emph{K} selected segments intact, and impute the rest of the segments prior to classifying the clip. The outcome of this experiment is summarized in Table~\ref{table:cumulative_importance}.        


\begin{table} [!htb]
\vspace{-5pt}
\centering
\resizebox{\linewidth}{!}{
\begin{threeparttable}
\begin{tabular}{|l|l|rrrrr|} 
\hline
\textbf{Sound Category} & \textbf{\#Clips} & \textbf{K=10\%}  & \textbf{K=30\%} & \textbf{K=50\%} & \textbf{K=70\%} & \textbf{K=90\%}
\\
\hline
\textbf{Voice Command} & 600 & 64.20\% & 81.50\% & 88.70\%  & 100.00\% & 100.00\%
\\
\textbf{Vibration} & 120 & 56.62\% & 81.18\% & 92.31\% & 93.47\% & 94.20\%
\\
\textbf{Alarm} & 85 & 54.62\% & 85.09\% & 91.91\% & 93.11\% & 93.50\%
\\
\textbf{Children} & 80 & 56.62\% & 88.82\% & 92.19\% & 94.61\% & 95.95\%
\\
\textbf{Music} & 76 & 48.52\% & 82.29\% & 89.61\% & 91.47\% & 94.14\%
\\
\textbf{Pet} & 76 & 52.27\% & 83.61\% & 92.10\% & 94.27\% & 95.10\%
\\
\textbf{Animal} & 65 & 54.68\% & 90.63\% & 92.21\% & 93.00\% & 100.00\%
\\
\textbf{Bursts} & 16  & 51.13\% & 65.90\% & 81.81\% & 85.20\% & 92.00\% 
\\
\textbf{Laundry} & 16  & 54.10\% & 69.70\% & 83.10\% & 89.40\% & 98.00\% 
\\
\textbf{Sleep} & 16  & 56.21\% & 72.27\% & 80.53\% & 88.96\% & 100.00\%
\\
\textbf{Wellness} & 16  & 64.32\% & 85.42\% & 88.78\% & 100.00\% & 100.00\%
\\
\textbf{Bathroom} & 12  & 61.11\% & 83.33\% & 90.00\% & 100.00\% & 100.00\%
\\
\textbf{Drink} & 12  & 55.32\% & 67.40\% & 84.18\% & 86.24\% & 95.00\% 
\\
\hline
\textbf{Overall} & 1190 & 51.47\% & 85.91\% & 89.60\%  & 92.30\% & 96.60\%
\\

\hline
\end{tabular}
\captionsetup{font=Large}
\caption{The percentage of clips that are classified correctly when we sample \emph{K\%} most informative segments.}
\label{table:cumulative_importance}
\begin{tablenotes}[para,flushleft]
\end{tablenotes}
\end{threeparttable}
}
\vspace{-13pt}
\end{table}

\begin{figure}[!thb]
  \centering
  \begin{subfigure}[b]{0.225\textwidth}
    \includegraphics[width=\textwidth]{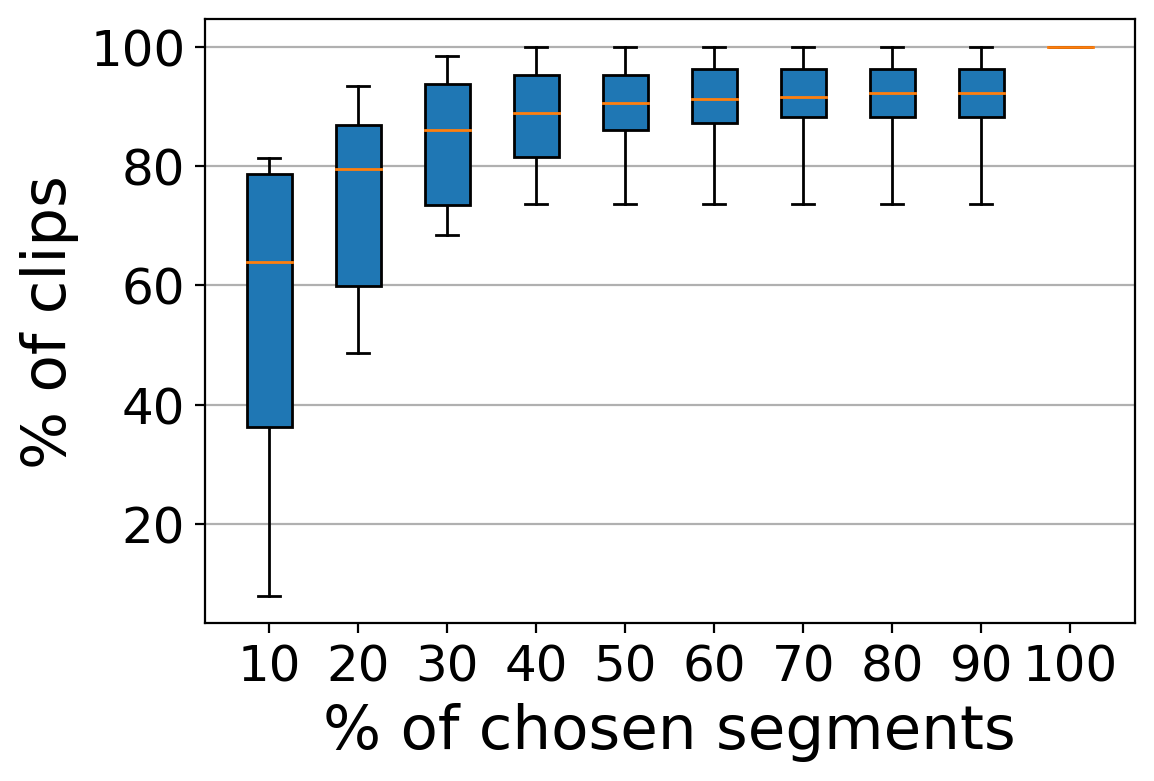}
    \caption{Urban8k}
  \end{subfigure}
  \hfill
  \begin{subfigure}[b]{0.225\textwidth}
    \includegraphics[width=\textwidth]{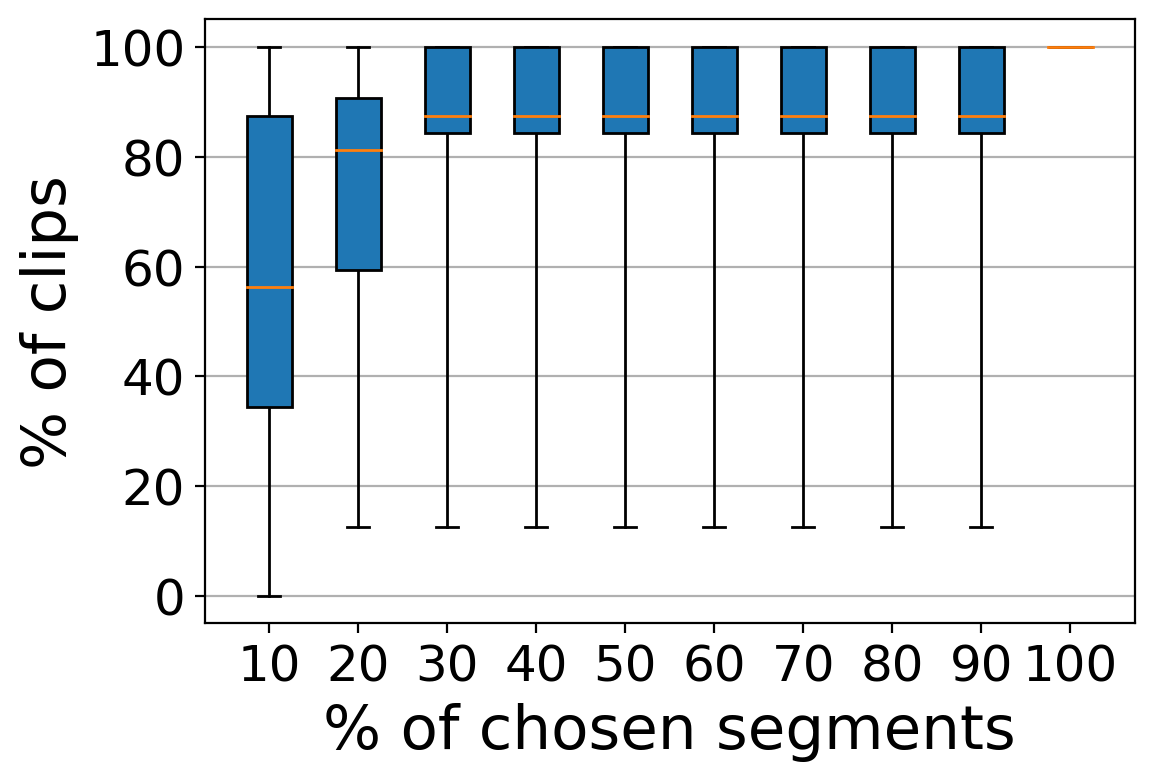}
    \caption{ESC-indoor}
  \end{subfigure}
  \\
  \begin{subfigure}[b]{0.225\textwidth}
    \includegraphics[width=\textwidth]{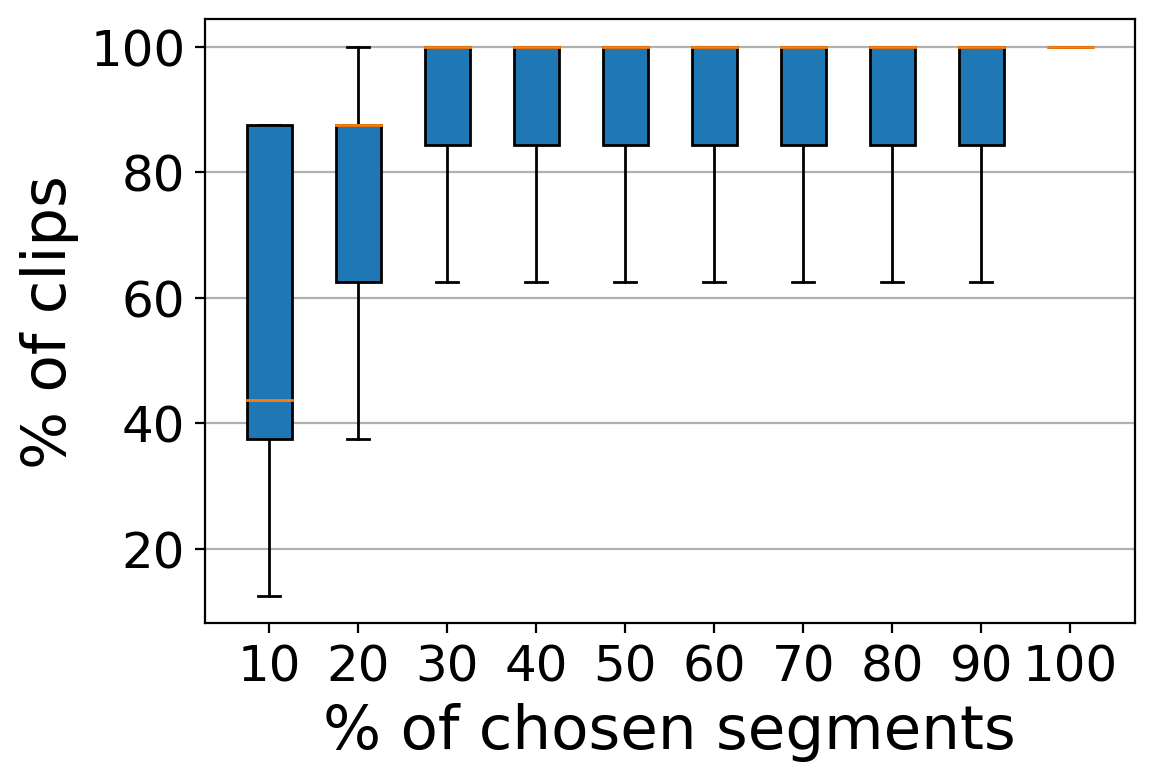}
    \caption{ESC-animal}
  \end{subfigure}
  \hfill
  \begin{subfigure}[b]{0.225\textwidth}
    \includegraphics[width=\textwidth]{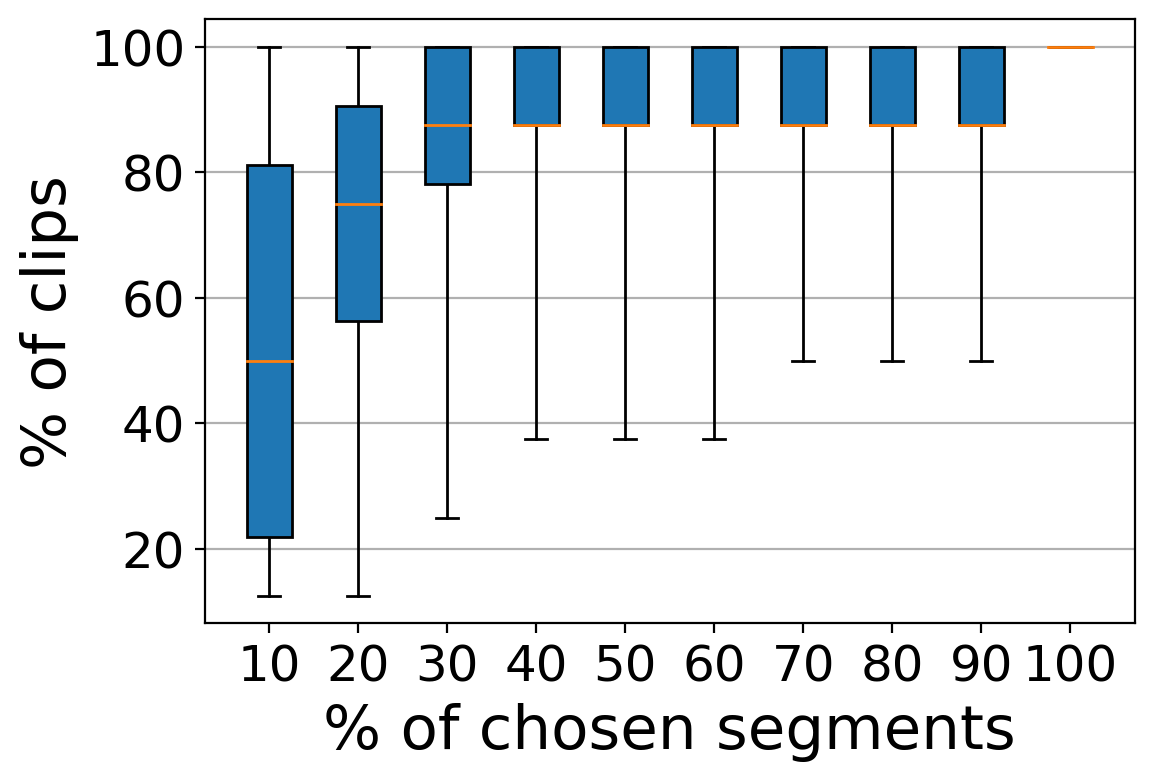}
    \caption{GSC}
  \end{subfigure}
  \caption{The distribution of correctly-classified clips in each dataset when \emph{K} most informative segments are sampled.}
  \vspace{-5pt}
  \label{fig:CDFs}
\end{figure}

Each row of Table~\ref{table:cumulative_importance} is a cumulative distribution of \emph{K} for a certain audio category, where \emph{K} denotes the minimum number of most informative segments to sample to classify an audio correctly. For example, the last number of the first row tells us -- if we sample the most informative 90\% segments (and impute the remaining 10\%), we can correctly classify 93.5\% of the alarm sounds from Urban8k dataset. Figure~\ref{fig:CDFs} shows the same information as in Table~\ref{table:cumulative_importance} but separately for each dataset.      

This result provides insights into the redundancy in audio clips for classification purposes. For an intermittent system, from its charge and discharge times, we can estimate what fraction of the time the system can sample, and then estimate their expected inference accuracy from these distributions. These datasets, however, are trimmed and preprocessed to remove unwanted sounds. In real-world audio, we expect more redundancy and thus an intermittent system is expected to have more room for selecting informative segments to achieve higher accuracy than the estimated values in this study. In conclusion, \textbf{not all audio segments being equally informative, a content-aware sampler on a resource-constrained sensing system can select a subset of informative audio segments to maximize their ability to correctly classify long-duration audio clips.}   

\subsection{Sparsity of Audio Events}

Using a large capacitor for intermittent audio event classification ensures continuous energy supply, but can result in missed events if charging time is longer than gaps between consecutive events. 

\begin{figure}[!htb]
    \centering
    \vspace{-5pt}
    \includegraphics[width=0.9\linewidth]{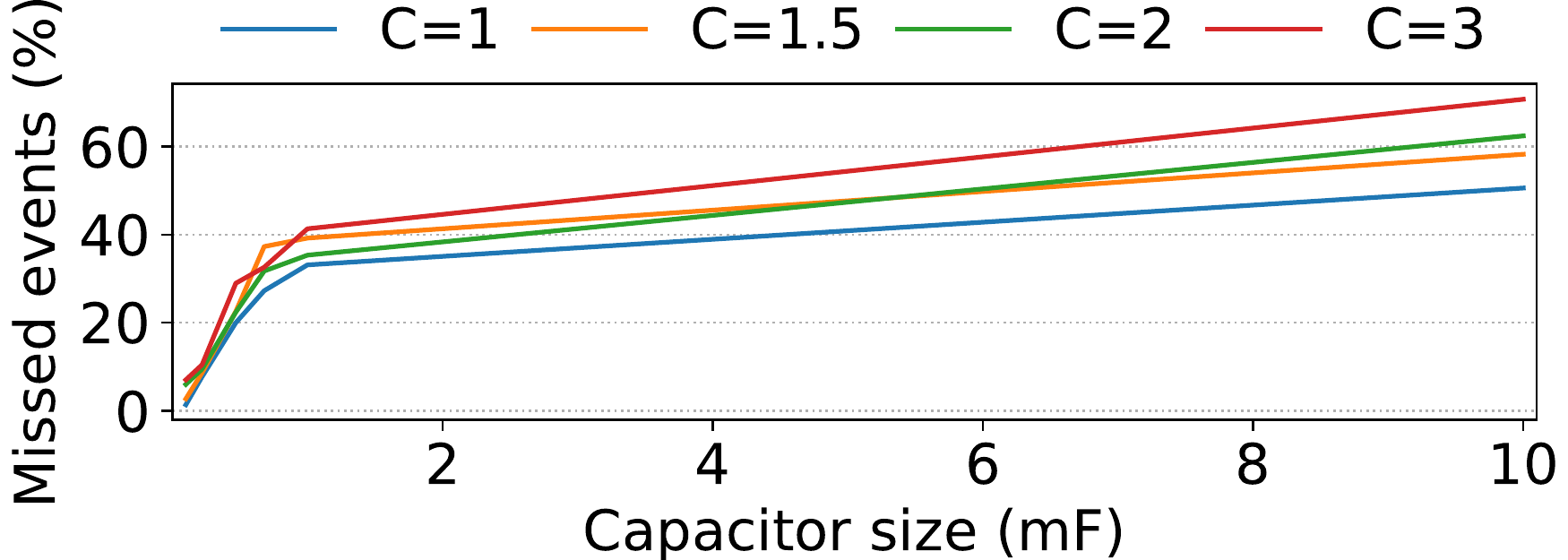}
    \caption{Percentage of audio events that are missed for different capacitor sizes at different energy harvesting rates. C denotes the ratio between charging and discharging times of the capacitor.}
    \label{fig:motivation_1}
\end{figure}

We conduct an experiment using the DCASE SELD dataset~\cite{DCASE_SELD} to measure the percentage of missed audio events due to capacitor charging and discharging times in intermittent systems with varying capacitor sizes ($100 \mu F$ to $10mF$) and energy harvesting parameters ($1$ to $3$). The dataset contained $600$ audio scenes, recorded in different environments with real-world background noise. The results are used to quantify what fraction of the audio events are missed (completely) by a typical intermittent system.


Figure~\ref{fig:motivation_1} shows that with larger capacitors, e.g., $10mF$, as many as 70\% of the audio events are missed in real-world scenarios. In contrast, smaller capacitors, e.g., $100\mu F$, results in as low as 4\% missed audio events, but the captured audio events have many missing samples or holes in them -- which calls for an intelligent audio sampling and processing technique that deals with missing samples in intermittent audio sensing systems.



\begin{figure*}[!thb] 
    \centering
    \subfloat[Offline Phase]{
        \includegraphics[width=0.5\textwidth]{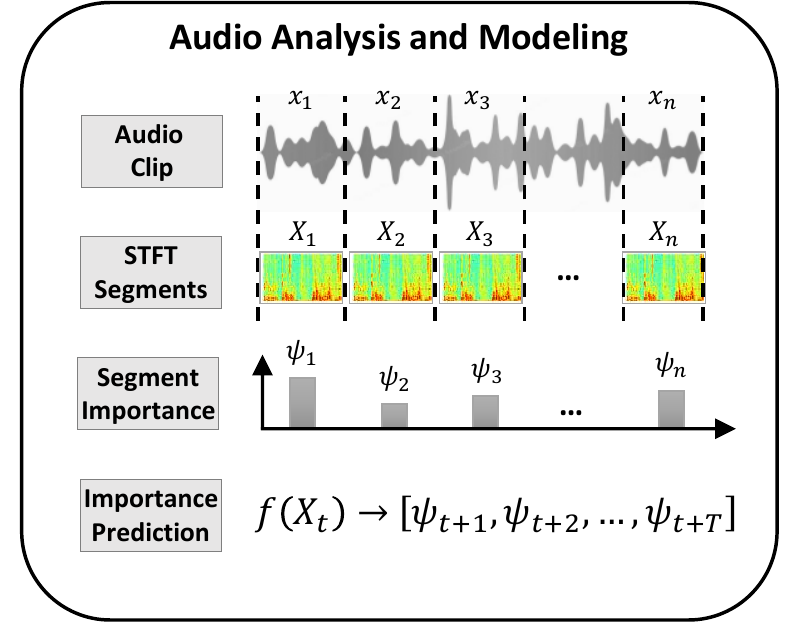} 
    }
    \subfloat[Online Phase]{
        \includegraphics[width = 0.5\textwidth]{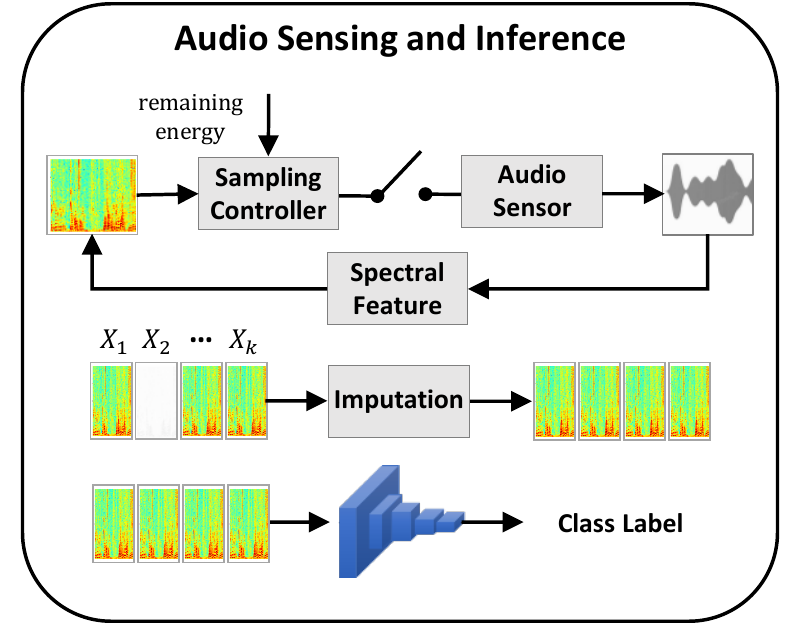}     
    }
    \caption{\sys consists of two phases: (a) an offline phase for audio analysis and modeling, and (b) an online (real-time) phase for audio sensing and inference on an intermittent system. }
    \label{fig:overview}
\end{figure*} 


\section{Overview of \SYS}

\sys is an audio sensing and inference framework that enables seconds-long audio event recognition on intermittently-powered systems. At the heart of \sys is a lightweight, content-aware sampling algorithm that decides (in real-time) when the system should wake up from sleep to actively sample the next few segments of the incoming audio. \sys's sampling objective is to maximize the classification accuracy by capturing the most informative audio segments, given the remaining charge in the capacitor, which limits the maximum number of segments the system can sample before it runs out of harvested energy.  

\sys is a software-only solution which does not require specialized hardware or multiple collaborating nodes. The framework consists of two phases: (a) an offline phase for audio analysis and modeling, and (b) an online (real-time) phase for audio sensing and inference on an intermittent system. The offline phase involves processing a large dataset of audio recordings to extract features and train machine learning models for both audio event sampling and classification. The online phase involves deploying the trained models to an intermittent system, where the system can detect audio events in real-time. The online phase operates on a continuous cycle of energy harvesting, sensing, and classification. Figure~\ref{fig:overview} shows both phases along with the major processing steps in them.                  




\vspace{-5pt}
\subsection{Audio Analysis and Modeling}

\parlabel{Audio Segment Importance.} \sys logically divides audio clips into non-overlapping 100ms \emph{segments}, and processes one segment at a time. The segment size is empirically determined to optimize the system's ability to extract acoustic features under different energy harvesting conditions. Each audio segment is assigned an importance score that denotes how important that segment is for accurate classification of the clip. During the offline phase of \sys, the importance score of each 100ms segment of every audio clip in the dataset is computed.     



\parlabel{Predicting Segment Importance Scores.} In order for \sys to decide which segments to capture next, the system must be able to predict the importance of the next few segments in the incoming audio. To enable this, during the offline phase of \sys, a simple regression network (which runs on the intermittent platform in real-time) is trained. The predictor predicts the importance score of up to $5$ next segments, given the frequency domain features of the audio segment that has just been read.


\vspace{-5pt}
\subsection{Audio Sensing and Inference}

\parlabel{Sampling.} \sys's audio processing system is triggered by the microphone, which remains in ultra-low-power listening mode and sends an interrupt to the microcontroller whenever a sound is detected. \sys reads an audio segment and extracts its spectral features. The sampling controller uses these features and the offline-trained importance predictor to identify potential candidate segments to read, and based on the remaining charge in the energy buffer, it decides whether (and for how long) it should go into to the low-power mode before it samples the next segment.        




\parlabel{Imputation.} Once the audio event ends, prior to classifying, the missing segments in the sampled audio are imputed to estimate the missing values. A lightweight, hierarchical interpolation technique is used to impute the missing segments. Note that even though the missing segments are less important (according to the segment importance score predictor), \sys still imputes them to form a continuous audio clip which can be fed to a neural network.


\parlabel{Classification.} An offline-trained convolutional neural network (CNN) is used to classify the audio. The classifier is trained on an augmented dataset where missing values are synthetically introduced at random locations in the audio clip and then the audio is imputed. To be able to classify audio signals of arbitrary lengths, a global max pooling layer is used after the final convolution layer which sums out the spatial information, making it robust to the arbitrary spatial dimension of the input. 

\section{Audio Analysis and Modeling}
\label{sec:offline}

During the offline phase, \sys analyzes each audio clip in the dataset to estimate the importance of each segment and to generate a predictor that is able to predict the importance scores.    


\subsection{Audio Segment Importance}
\label{sec:localscore}

Since harvested energy is scarce, \sys aims to sample only the highly informative segments to make the best use of the energy. Hence, there has to be an objective measure that quantifies the importance of each segment of an audio clip -- which the sampling controller can use in its segment selection process.        

\parlabel{Approach.} A naive way to identify important segments of an audio clip is to remove that segment, estimate its value using the remaining segments, and then test if this modified audio clip can still be classified correctly by a classifier that correctly classifies the unmodified original clip. This technique, however, does not provide a score that quantifies the importance of the segment.      

\sys determines the importance of each segment of an audio clip by analyzing its relative contribution towards the classification of the clip. The idea is similar to feature selection problem when each segment is treated as a feature. A linear regressor takes these features as the input and tries to predict the output -- which is a value as predicted by a baseline classifier (e.g., a neural network). The weights of the linear regressor indicates the relative importance of each feature (i.e. segment) when it tries to mimic the behavior of the baseline classifier.          

\parlabel{Algorithmic Details.} Given, a baseline classifier, $f$, that is trained on the entire dataset; and an audio clip, $\textbf{x}$, consisting of $n$ segments, following algorithmic steps are followed to quantify the importance of each segment:

\textbf{Step 1 --} Generate a large number of audio clips having missing values, $\textbf{z}_i$, by randomly turning on and off arbitrary number of segments in $\textbf{x}$.   

\textbf{Step 2 --} Impute all missing values in $\textbf{z}_i$ using a fast, lightweight imputation algorithm described later in this paper (Section~\ref{sec:imputation}).    

\textbf{Step 3 --} Train a linear ridge regressor, $g$ that minimizes a measure of how unfaithful $g$ is in approximating $f$ in the locality defined by a cosine distance metric. The loss function is as follows:
\[
    \argmin_{\textbf{w}} \sum_{i} d(\textbf{x}, \textbf{z}_i) \big[f(\textbf{z}_i)-g(\textbf{z}_i; \textbf{w})\big]^2
\]
Here, $d(\textbf{x}, \textbf{z}_i)$ is a distance measure between the original clip, $\textbf{x}$ and the augmented clip, $\textbf{z}_i$; $\textbf{w}$ is the learned weight vector whose elements $\{w_k\}$ are the desired importance scores of the segments in $\textbf{x}$.    

These three steps described above are repeated for each audio clip in the dataset.

\begin{figure}
    \centering
    \includegraphics[height=1.5in]{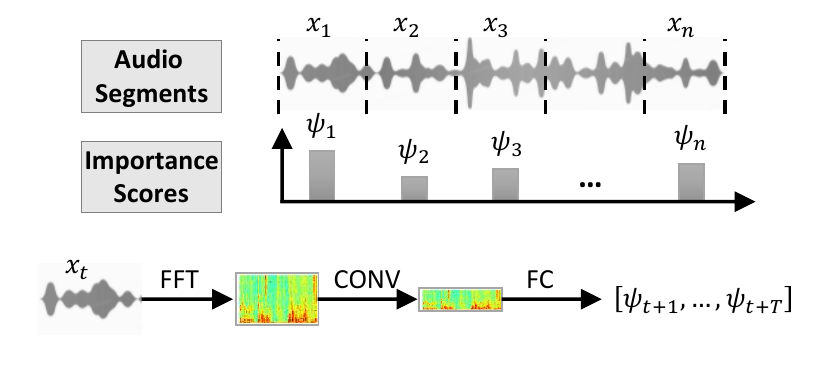}
    \caption{Segment Importance Predictor.}
    \vspace{-20pt}
    \label{fig:scorepredict}
\end{figure}

\subsection{Predicting Segment Importance Scores}

Since the sampling controller in \sys has to predict the importance of the next few segments that are yet to be sampled, it requires a predictor that can do so from the knowledge of recently sampled audio. As this predictor has to run on the intermittent system in real-time, it must be lightweight and low-overhead.    


\parlabel{Approach.} Given an observed audio segment at any instant, \sys employs a lightweight neural network to predict the importance scores of the next few segments of the audio. The network is trained during the offline phase and used in the online (real-time) phase of \sys.

The network consists of fully connected layers -- which makes it fast and low-overhead. The network, however, takes spectral features of the audio segment as the input since they carry unique signatures of various types of audio events. Even though the predictor uses spectral features, these features are identical to the first layer of the audio classier (described in Section~\ref{sec:classification}) which are reused. This is why they do not add extra overhead to the end-to-end audio processing cost in \sys.       

\parlabel{Algorithmic Details.} The development and training process of the segment score predictor is as follows:            


\textbf{Step 1 -- } For each audio clip in the dataset, we randomly take a segment from the clip and take the importance scores of the next 5 segments. Depending on the length of the audio, the number of segments in a clip can be anything. \sys's goal is to predict the next $5$ segments’ importance scores, given the current segment. To train this predictor, from each $n$-segment clip, $(n-6+1)$ training examples are generated.


\textbf{Step 2 --} Segments chosen in the previous step undergo a feature extraction process where a 1x4 convolution operation is applied after taking the short-time Fourier transform of the segment.       

\textbf{Step 3 --} The regression neural network is trained using the spectral features as the input and the importance scores as the output.   




\parlabel{Rationale behind Convolution.} The convolution layer has a filter of dimension (1x4); hence, the filter only convolves along the frequency axis but not the time axis. The reason behind this design is three-fold. First, a filter that convolves along the time domain is not feasible since a segment is observed only for a very short period in time. Second, due to memory constraints, the dimensions of the sampled data needs to be lowered, which is achieved by convolution. Third, by using a convolution layer which overlaps with the audio classifier -- we are able to extract fine-grained information for predicting the importance of next few segments.


\subsection{Global Segment Importance}
After computing the \emph{local} importance of each segment for each audio clip, we aggregate the scores to compute a \emph{global} importance score of the audio segments over the entire dataset. The process of computing global importance score of the audio segments is as follows:

\textbf{Step 1 -- } The importance score of each segment of each clip are normalized to the range [-1, +1]. 

\textbf{Step 2 -- } The average of all importance scores for each segment over all audio clips in the whole dataset is computed. This gives us a global importance score of each audio segment for the entire dataset. 

Both global and local importance scores are used by \sys to decide which segment to sample next. The rationale behind having a global importance score for each segment is that when none of the next few segments have significantly high local importance scores or there is a lot of uncertainty in the prediction of the local importance scores, but the system has energy to sample one or more segments, it needs another means to be able to rank the segments. This technique complements the local score-based sampling and improves \sys's ability to sample informative segments.

\section{Audio Sensing and Inference}


During the online (real-time) phase, \sys intelligently samples audio segments, imputes the missing values, and performs on-device inference.

\subsection{Sensing and Preprocessing}

\sys remains in its ultra-low-power sleep mode when there is no sound in the environment and it has no pending tasks. This is enabled by the ultra-low-power listening mode of the microphone which continuously listens to the environment while being powered by harvested energy and sends an interrupt signal to the microcontroller only when audio activities are detected in the environment. This is when the microcontroller wakes up and switches to active mode (subject to availability of adequate harvested energy) to start its real-time audio sensing and inference processes.     

\sys samples audio data at the unit of 100ms segments. After reading an audio segment, it analyzes the segment (as part of the sampling controller algorithm that is described next) to decide whether to read the next segment or to go to sleep. This analysis process involves computing the spectral features of the audio segment that takes significant amount of time. Hence, \sys samples and pre-processes a segment in parallel, so that sampling decisions can be made in real-time. 

This is implemented by maintaining a four-element FIFO buffer queue, where each buffer holds 25ms of audio. The key idea is to pre-process data from a subset of the buffers in the low-energy accelerator (LEA) co-processor while the other buffer fills in with new data. The steps are as follows:

\textbf{Step 1 --} Initially, the first two buffers fill up (25ms + 25ms = 50ms). 

\textbf{Step 2 --} While the spectral features are extracted on buffers 1 and 2, in parallel, the third buffer fills up.   

\textbf{Step 3 --} While the spectral features are extracted on buffers 2 and 3, in parallel, the fourth buffer fills up.    

\textbf{Step 4 --} The spectral features are extracted on buffers 3 and 4. 

By using these buffers, we compute FFT of 50ms audio, which means the window length for the STFT is 50ms.
We are also moving our window by one buffer during each iteration, meaning we are using a hop length of 25 ms. Additionally we are also using a convolution layer of shape (1x4), which means the convolution is done only in the frequency axis and not in the time axis. Hence it is not dependent on the previous or next temporal information.



\vspace{-10pt}
\subsection{Sampling Strategy}

\sys makes sampling decisions at runtime based on both the global as well as the local importance scores of the segments. The global importance scores are used to have an initial plan for segment selection and sampling. This initial sampling plan is modified as the system acquires new segments and gets more insights into the importance of the segments based on their local importance scores.   

\parlabel{Parameter C.} \sys keeps track of its energy harvesting rate by computing the ratio, $C$ between the charging and discharging times of the capacitor. This ratio characterizes the dynamics of an intermittent system. For example, a continuously-powered system has $C = 0$, while any intermittent system has $C > 0$. If $C = 1$, an intermittent system has to charge for the duration of exactly one segment in order to fully recover the energy it lost by sensing one segment. Therefore, the value of $C = 1.0, 1.5, 2.0, 3.0$ refers to $50\%, 60\%, 67\%, 75\%$ missing values in the data, respectively. Generally, $C = 1$ represents outdoor light (1200 lux), $C = 2$ represents indoor light (800 lux), and $C = 3$ represents dimly lit room (500 lux) and weak RF sources. Recent works on intermittent sensing have used a value of $C$ within similar ranges. For example,~\cite{continuous} used $1 < C < 3$, and~\cite{fulvio} used $0.8 < C < 1$ (outdoor solar), $1.1 < C < 3$ (indoor solar), and $1.0 < C < 3.2$ (indoor RF).

\parlabel{Initial Sampling Strategy.} \sys uses the parameter $C$ to formulate an initial sampling strategy as follows: 

\textbf{Step 1 --} At first the maximum number of segments, $t_{max}$ that can be sampled is estimated from the current charging-to-discharging ratio, $C$. For instance, if $C = 2$, the system has to wait twice the amount amount of time in low-power sleep mode to recharge the capacitor to gain back the same amount of energy that is used for sampling and processing audio. This sets a limit on the number of segments that the system can sample and use in classifying the audio event.

\textbf{Step 2 --} A binary mask is created that tells the system which segment to sample and which to skip. Using both the charge-to-discharge ratio, $C$ and the global segment importance scores, the mask values for the most important $t_{max}$ segments are set to 1. These are the segments that the system should sense in order to maximize its classification performance. 

\textbf{Step 3 --} In order to account for energy constraints, the mask values (i.e., the current sampling strategy) are checked to ensure that in no point in time the stored energy is completely exhausted, and if so, the mask value for that position is set to 0. This is a situation when the system must stop sampling for lack of energy even though the segment is important.     



\parlabel{Adapting the Initial Sampling Strategy.} \sys relies on the low-power listening mode of the microphone to remain in low-power mode waiting for an audio event to occur. Once the microphone interrupts \sys, the system transitions from low-power mode to active mode, and starts sampling and processing audio segments of length, $t_{segment}$. 

The system starts with the initial sampling strategy, but the plan changes as it encounters new audio segments. The process is as follows:   

\textbf{Step 1 --} STFT of the audio segment is computed and a 1x4 convolution is performed on the frequency axis to extract the frequency domain features. 

\textbf{Step 2 --} The segment importance predictor is used to predict the importance of the next $n$ segments. There are two cases: first, if the importance scores of any segment is higher than an empirically obtained threshold, or if the system is sitting idle with full charge, it decides to sample that segment that is locally important but may not have been included in the initial sampling plan. Second, if none of the importance scores is higher than the threshold, the system follows the initial sampling plan as the default.     

After deciding which segment to sample next, unless it is the very next segment, \sys switches to low-power sleep mode and wakes itself up when that segment arrives. This process is repeated until the end of audio event.
\subsection{Imputation}
\label{sec:imputation}

Although \sys's intelligent sampling algorithm has control over which segments to sample, it cannot avoid missing segments in the captured audio. Prior to classifying an audio clip that has missing values, \sys estimates the missing values using a frequency domain imputation technique.  

\parlabel{Approach.} Prior to imputation, \sys transforms time domain audio signals into frequency domain signals using the fast Fourier transform (FFT). The rationale behind imputing signals in the frequency domain is that audio events are characterized by their frequency components and the unique characteristics of different types of audio events are easily observable in the frequency domain. Furthermore, the frequency components of an audio event typically do not change drastically within a short period. Hence, interpolating the signal in the frequency domain is much more effective than doing so in the time domain.

\parlabel{Algorithmic Details.} Given a sequence of audio segments where some of the segments are null (missing valued), following steps are performed to estimate the missing values:

\textbf{Step 1 --} Suppose, there are missing segments from time step, $t_s$ to time step, $t_e$. Hence, the length of the missing segments is $t_s-t_e+1$. The imputation process starts from these two ends, $t_s$ and $t_e$. 

\textbf{Step 2 --} At each time step, $t$, the frequency domain components are interpolated using following equations:

\begin{equation}
    X(t,f)=\big[1-r(t)\big]X(t_s-1,f) + r(t)X(t_e+1,f)
    \label{eq2}
\end{equation}
\begin{equation}
    r(t)=\frac{t-(t_s-1)}{(t_e+1)-(t_s-1)}
    \label{eq1}
\end{equation}

\begin{figure}[!htb]
    \centering
    \includegraphics[width=\linewidth]{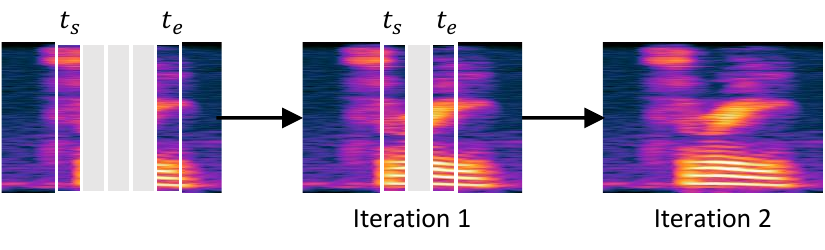}
    \caption{Illustration of imputation: during the first iteration, audio segments at time $t_s$ and $t_e$ are interpolated, leaving only one segment left for imputation. The process is repeated until all segments are imputed.}
    \label{fig:imputation}
\end{figure}

\textbf{Step 3 --} The previous step is repeated after the value of $t_s$ has been increased and the value of $t_e$ has been decreased, until $t_s < t_e$.  

\subsection{Inference}
\label{sec:classification}

\sys employs a convolutional neural network (CNN) to classify the audio event. The CNN consists of a combination of convolution, max pooling, and fully connected layers. The network architecture is shown in Table~\ref{tbl:netarch}. 

\begin{table}[!htb]
\centering
\resizebox{.5\columnwidth}{!}{\begin{tabular}{|l|l|l|l|}
\hline
 \textbf{\#} & \textbf{Layer} & \textbf{Dimensions} & \textbf{Parameters}\\
 \hline
  1 & Conv2D & 60x32x1 & 5\\
  2 & Conv2D & 29x30x2 & 20\\ 
  3 & Conv2D & 12x13x8 & 152\\ 
  4 & Conv2D & 4x4x32 & 2336\\ 
  5 & Dense & 256 & 8448\\ 
  6 & Dense & 10 & 2056\\   
 \hline
\end{tabular}}
 \caption{Neural network architecture (13,017 parameters).}
\label{tbl:netarch}
\end{table}

\parlabel{Approach.} The inference process starts when sampling the audio event has completed as determined by a period of silence by the microphone in its ultra-low-power listening mode. 
    
After sampling has completed and the missing audio segments have been imputed, the resultant short-time Fourier transform (STFT) values are converted to Mel-frequency spectrogram to reduce the dimension of the input signal. The spectrogram passes through to a sequence of convolution and max pooling layers. After the last convolution layer, a global max pooling layer is used --- which takes the maximum value across time and frequency domains for each channel, and outputs one value for each of the channels. This design makes it possible for the model to handle audio signals of arbitrary lengths. Since audio events may have arbitrary lengths, a global max pooling layer is used to handle the variable-length input, instead of zero-padding. This makes the classifier model robust to audio events of arbitrary length.

The network is trained offline using both original audio clips as well as augmented audio clips where missing values are randomly introduced. The augmented audio clips are imputed using the method described in Section~\ref{sec:imputation}, before using them for training the network. This is done to ensure that the model is robust to missing audio segments.  


\section{Evaluation on Datasets}

\subsection{Experimental Setup}

\parlabel{Dataset.} We use two popular audio classification datasets that contain audio clips of several seconds containing environmental sounds. We also use two datasets containing short speech commands and phrases. We divided these datasets into multiple sub-classses depending on the type of the sound. For example, ESC-50~\cite{piczak2015esc} dataset covers a large variety of sound types such as animal sound, human activity, indoor noises etc. A brief summary of the datasets used in our study is shown in table~\ref{tbl:dataset}. The duration of these clips are 1--5 seconds. Since we want to demonstrate \sys's ability to classify multi-second audio, we collected some of our own data by recording as well as from online sources. Furthermore, we divided the dataset according to the sound types. 

\begin{table}[!htb]
\resizebox{\columnwidth}{!}{\begin{tabular}{|l|l|l|l|}
\hline
 \textbf{Dataset} & \textbf{Clips} & \textbf{Classes} &\textbf{Examples}\\
 \hline
 Urbank8k~\cite{urban} & 8,732 & 10 & Human, nature, mechanical.\\
 ESC-50~\cite{piczak2015esc} &  2,000 & 50 & Human activity, animal, indoor\\
 GSC~\cite{warden2018speech} & 105,829 & 35 & Speech commands. \\
Fluent Speech~\cite{lugosch2019speech} & 23,132 & 31 & Short phrases. \\
 \hline
\end{tabular}}
 \caption{Datasets used in our study.}
\label{tbl:dataset}
\end{table}

\parlabel{Modeling Intermittence.} We use two parameters to model intermittence during our evaluation. First parameter is $B$ which denotes the maximum number of audio segments the system can process without power failure. In our experiments all the audio segments have a length of $100 ms$. We also use another parameter $C$ to denote the ratio between charging and discharging time to process one audio segment of $100 ms$. In this way, we discretize the energy pattern of an intermittent system. We start with an initial budget of $B$ and each time we read and process audio segment of $100 ms$, we decrease the value of $B$. Additionally, whenever the system skips an audio segment and waits in low-power mode, we increase the energy budget by $\frac{1}{C}$. However since $C$ can be floating point value, we consider only the lower bound while checking for available energy budget. Whenever the budget becomes zero, the system can not sense and process an audio segment, even if the importance of that segment is high.

\parlabel{Baseline Solutions.} We use three baseline solutions for comparison: Vanilla, Periodic, and CIS~\cite{bhatti2017harvos}. In the vanilla approach, we emulate how an intermittent system normally works, i.e., sensing until the energy buffer is completely depleted and then waiting for it to be filled again. The periodic sampler samples every $n$th audio
segment, where $n$ depends on the value of $C$. CIS~\cite{bhatti2017harvos} uses multiple sensor nodes to capture and classify $283$ms audio. We implement this approach by sensing 3 consecutive segments (3 $\times$ 100 = 300ms) and then waiting for the energy buffer to be full again. We denote this baseline as CIS1 in our experiments.

\parlabel{Evaluation Metrics.} We use inference accuracy to quantify the performance of the classifier. Accuracy refers to the portion of correctly classified instances on a dataset. We use detection of sound events for environmental sound classification dataset such as UrbanSound8k and ESC-50, and detection of certain phrases for GSC and Fluent Speech dataset. We also divide these dataset into multiple categories depending on the types of sound.    

\subsection{Comparison with the State-of-the-Art}\label{comparison}

\begin{figure}[!htb]
    \centering
    \includegraphics[width=\linewidth]{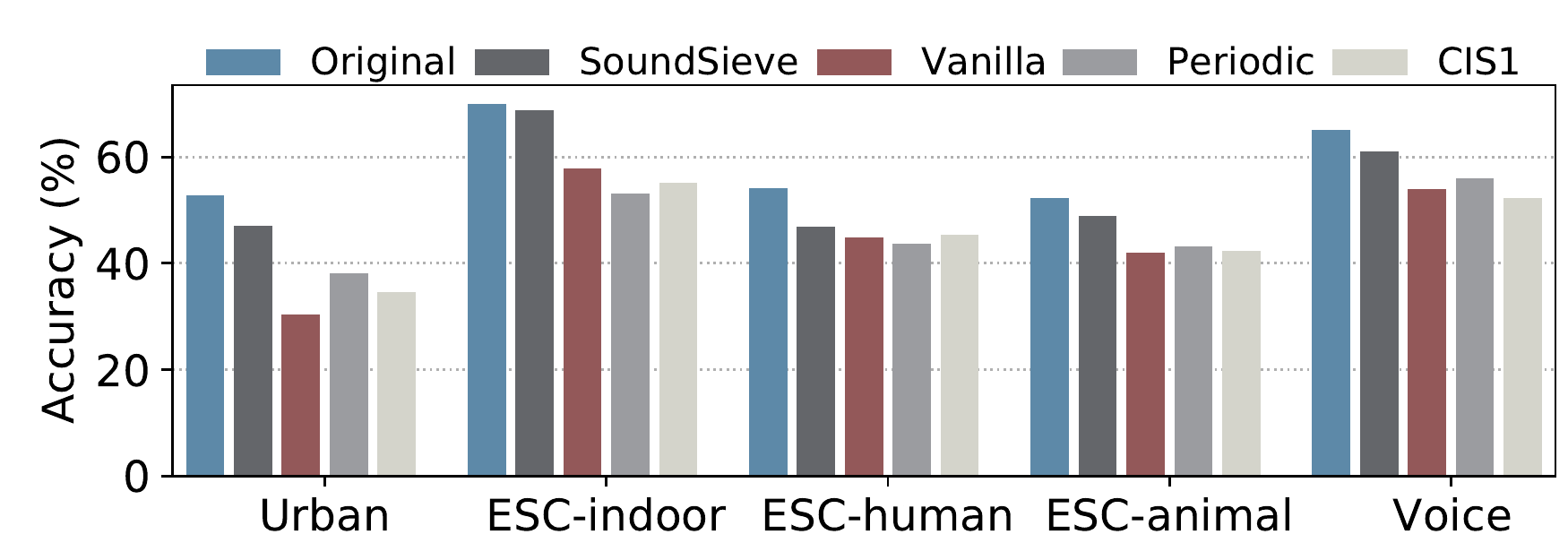}
    \caption{Comparison among different sampling approaches.}
    \label{fig:comp}
\end{figure}

In Figure~\ref{fig:comp}, we compare \sys's performance with that of state of the art approaches when the energy harvesting pattern is modeled by setting $C=1$. We also show the original accuracy of the classifier when there is no missing audio segment at all. Although the classification accuracy drops when there are missing audio segments due to power failure in an intermittent system, we see that conserving energy to sense when there are more information available improves the overall classification accuracy by 3\%-17\% than other approaches. 

\begin{figure}[!tbp]
  \centering
  \begin{subfigure}[b]{0.225\textwidth}
    \includegraphics[width=\textwidth]{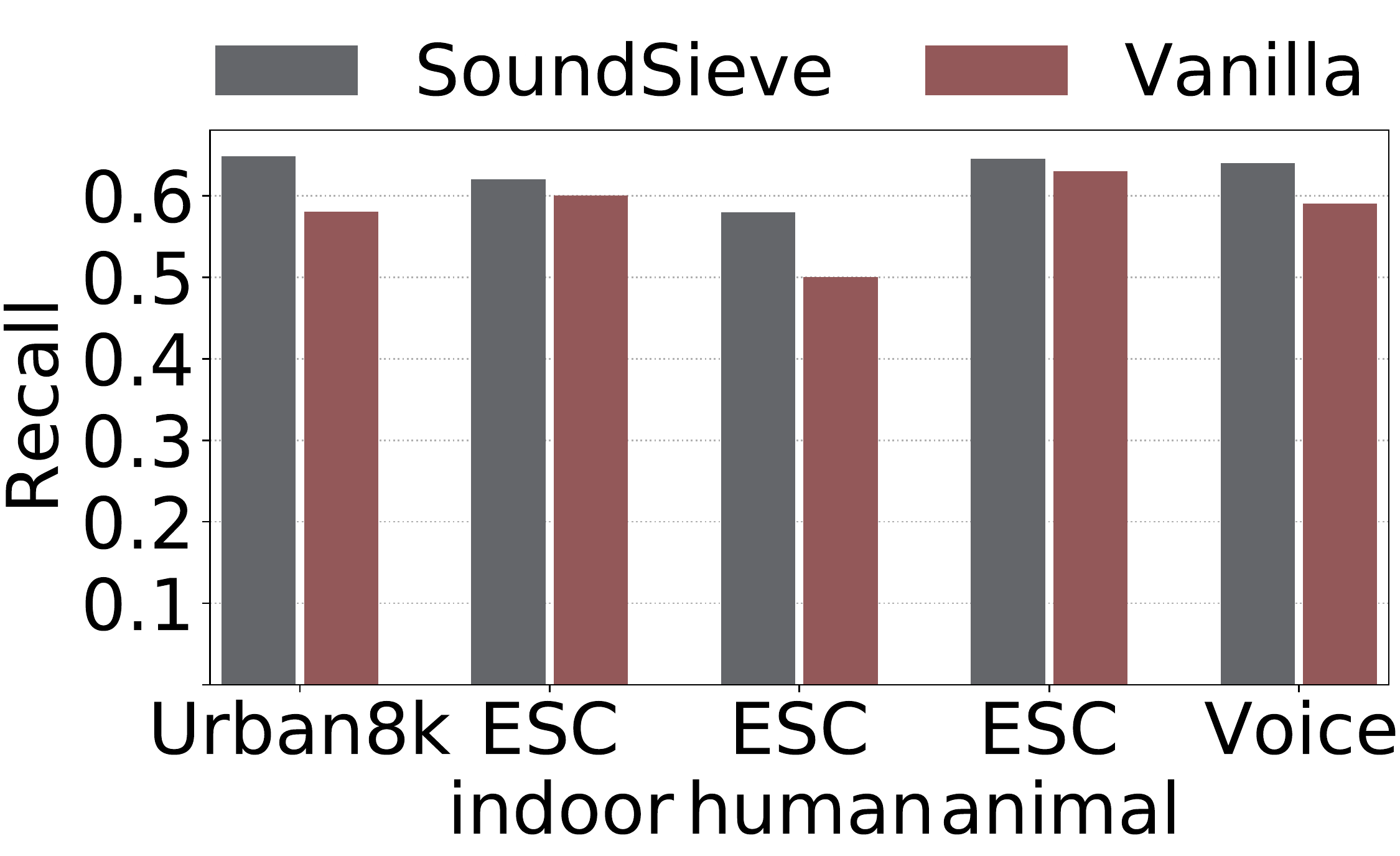}
    \caption{C=1}
  \end{subfigure}
  \hfill
  \begin{subfigure}[b]{0.225\textwidth}
    \includegraphics[width=\textwidth]{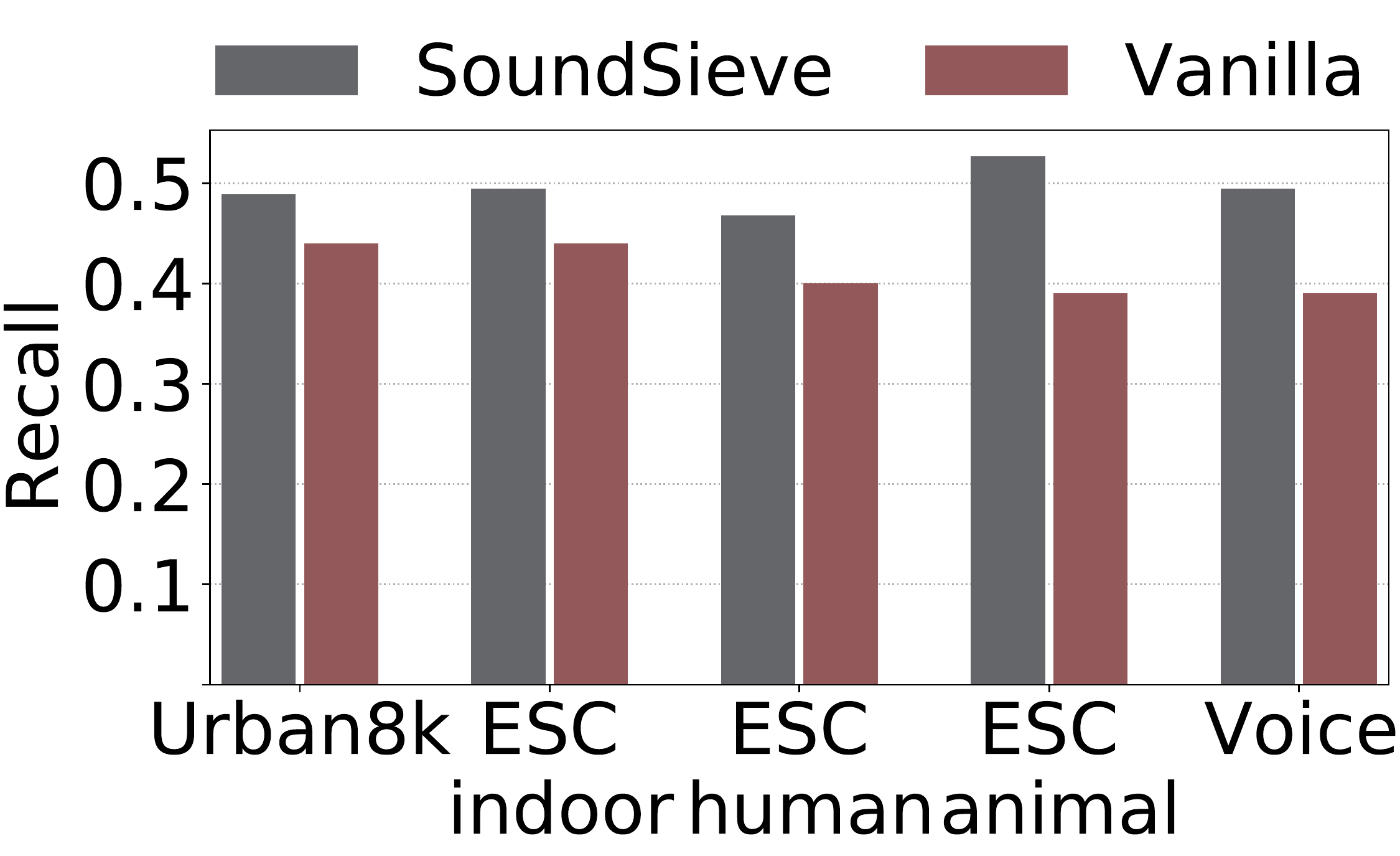}
    \caption{C=1.5}
  \end{subfigure}
  \\
  \begin{subfigure}[b]{0.225\textwidth}
    \includegraphics[width=\textwidth]{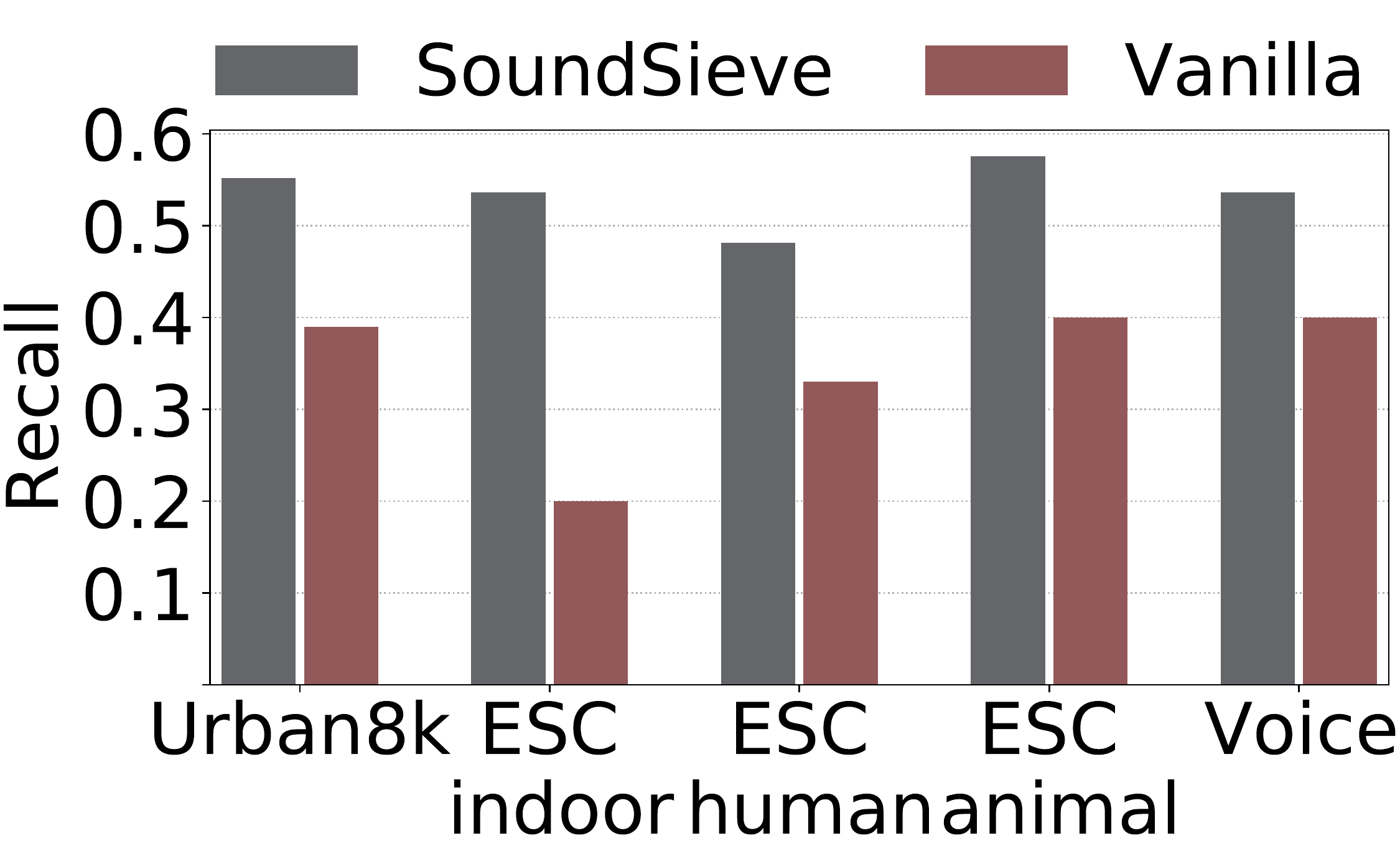}
    \caption{C=2}
  \end{subfigure}
  \hfill
  \begin{subfigure}[b]{0.225\textwidth}
    \includegraphics[width=\textwidth]{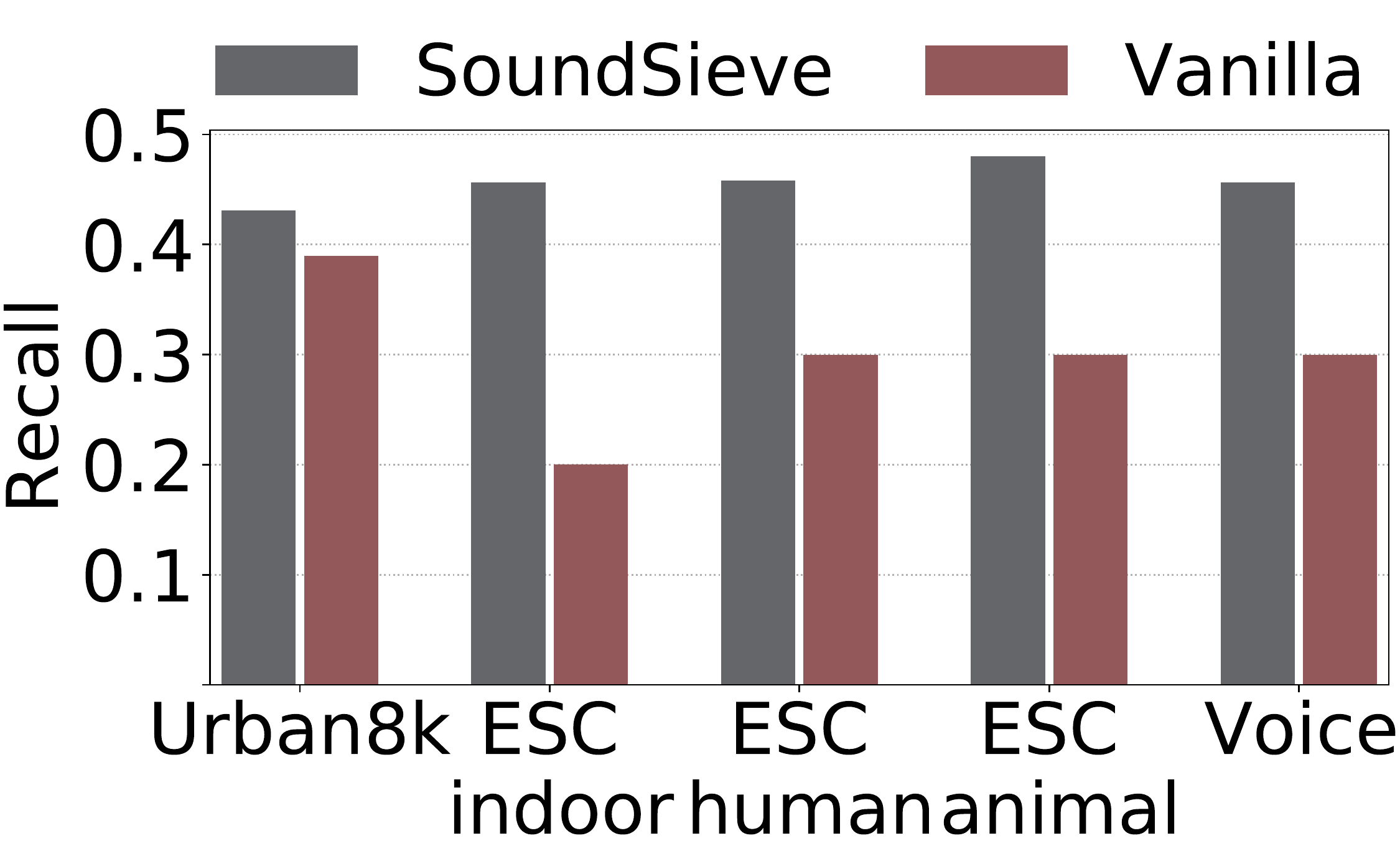}
    \caption{C=3}
  \end{subfigure}
  \caption{Comparison of coverage of the most informative audio segments at different energy harvesting rate}
  \label{fig:recall}
\end{figure}

\parlabel{Effectiveness of Importance Predictor.} In order to demonstrate why \sys can produce better classification accuracy than other approaches, we show the average coverage of the most informative segments in audio clips using both \sys and vanilla approach in Figure~\ref{fig:recall}. We use the term recall -- the ratio between the number of positive samples correctly labeled as positive to the total number of positive samples. 

\begin{equation}
    recall = \frac{true\ positive}{true\ positive + false\ negative}
\end{equation}

Here, a positive sample refers to an informative segment whereas a negative sample refers to a segment of lower importance. We use the true importance scores of each segment of an audio clip and the energy harvesting pattern to compute the most optimal sampling strategy and use that to measure the recall. 

The values of recall clearly demonstrate that \sys can sense as much as 25\% more important audio segments compared to the vanilla approach. As we decrease the energy harvesting rate by setting the value of $C$ higher, we observe greater improvement in covering only the most informative blocks compared to the vanilla approach. 

\subsection{Effect of Energy Harvester}
\begin{figure}[!htb]
    \centering
    \includegraphics[width=\linewidth]{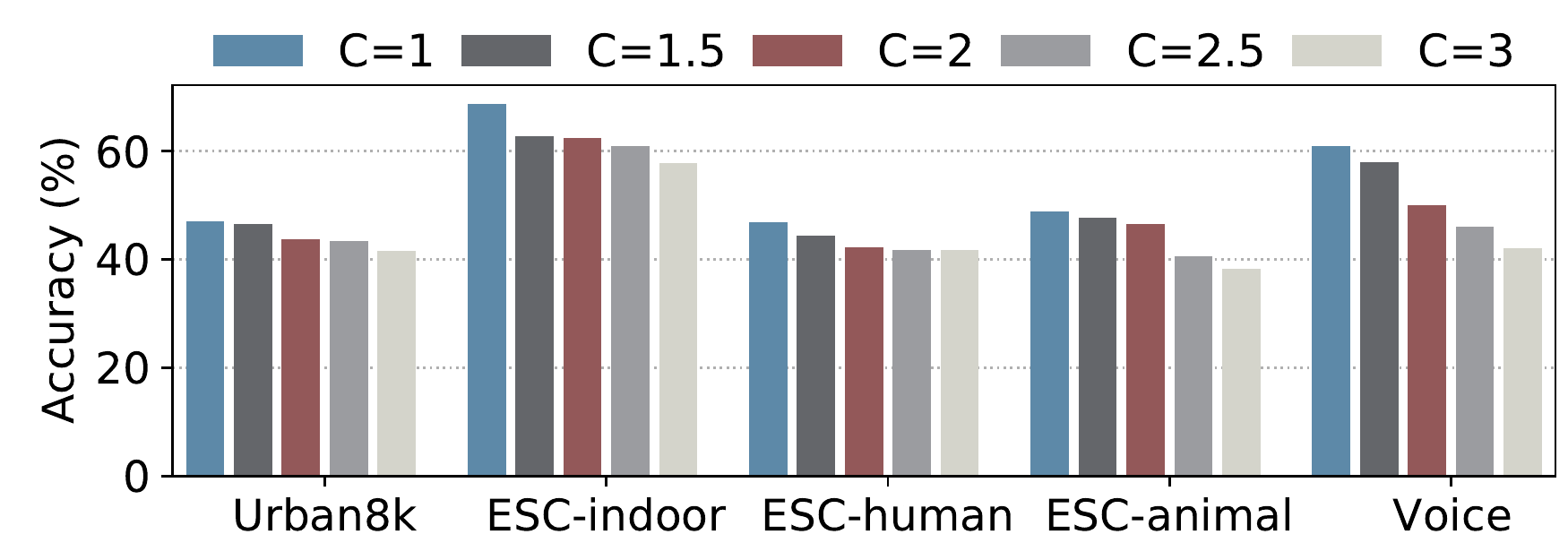}
    \caption{Effect of Harvesting Pattern}
    \label{fig:harverster}
\end{figure}

We change the energy harvesting pattern in our simulation by varying the value of $C$. A higher value of $C$ refers to a larger amount time required to harvest the energy of sensing one audio segment. It also means that a higher percentage of audio is missing in the audio clip. Figure~\ref{fig:harverster} demonstrates that as we decrease the energy harvesting rate, the classification accuracy drops slightly. However since \sys is able to cover more informative blocks as mentioned in~\ref{comparison}, the drop in classification performance due to change in energy harvesting pattern is minimal compared to other approaches.

\begin{figure}[!htb]
  \centering
  \begin{subfigure}[b]{0.225\textwidth}
    \includegraphics[width=\textwidth]{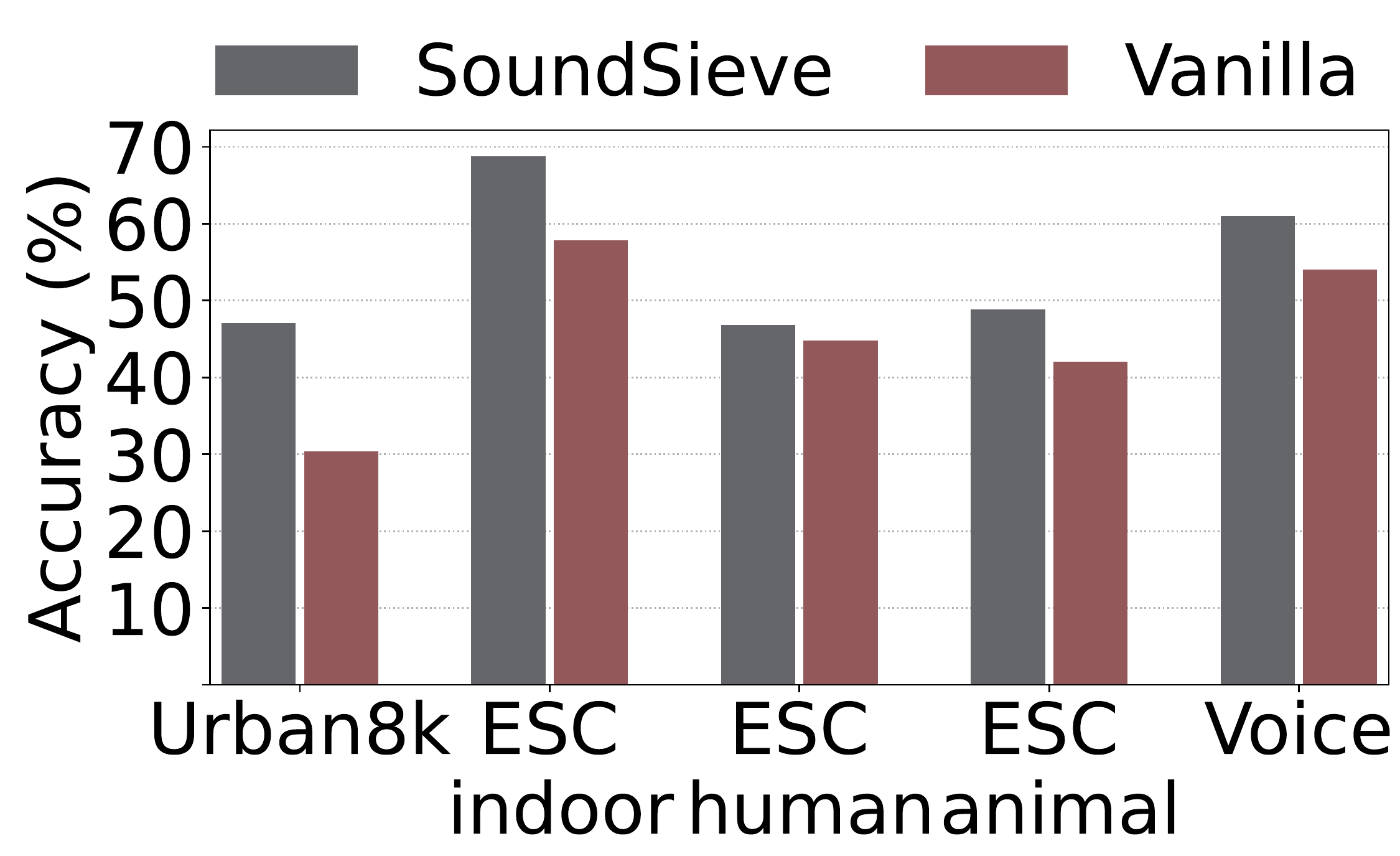}
    \caption{C=1}
  \end{subfigure}
  \hfill
  \begin{subfigure}[b]{0.225\textwidth}
    \includegraphics[width=\textwidth]{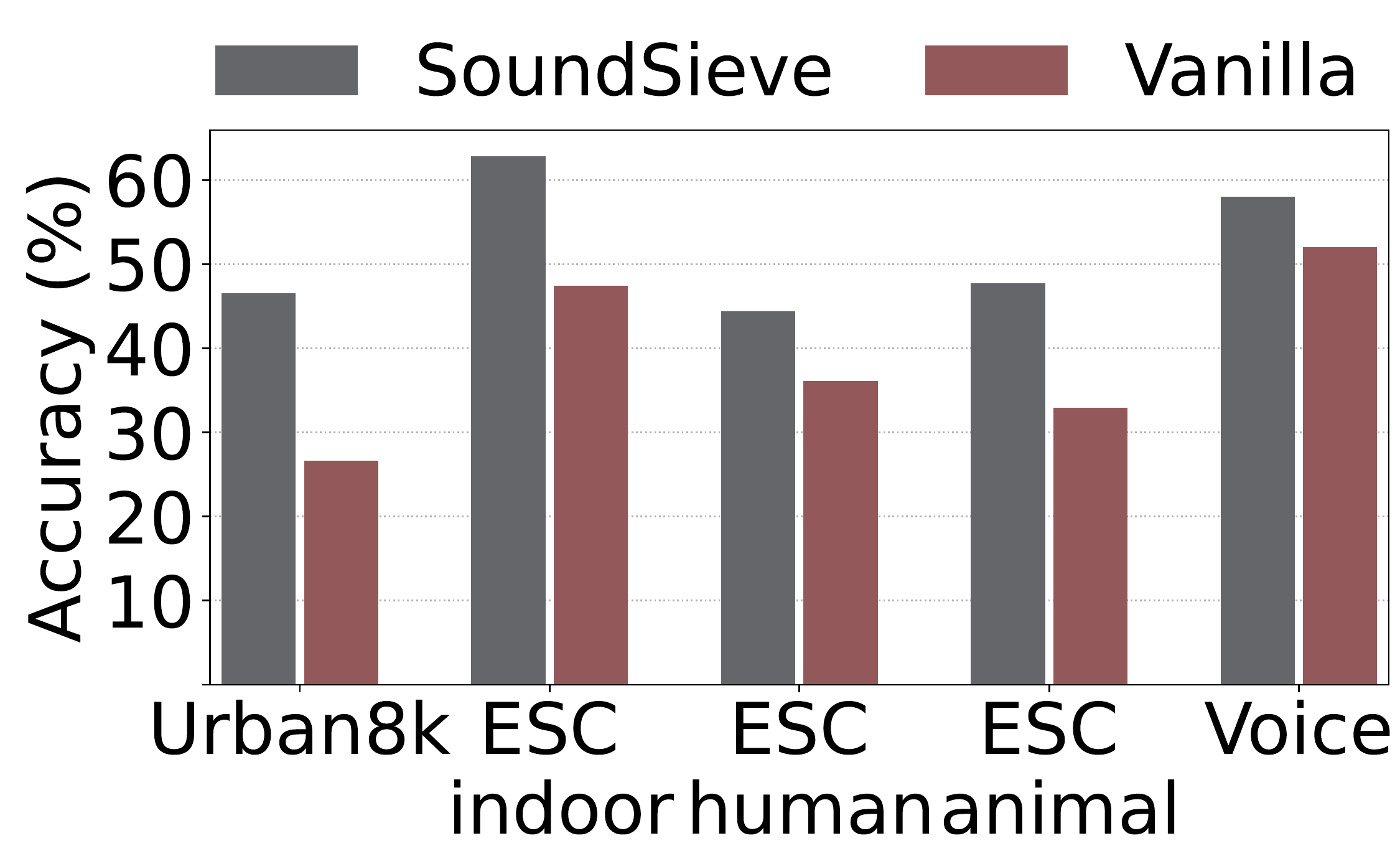}
    \caption{C=1.5}
  \end{subfigure}
  \\
  \begin{subfigure}[b]{0.225\textwidth}
    \includegraphics[width=\textwidth]{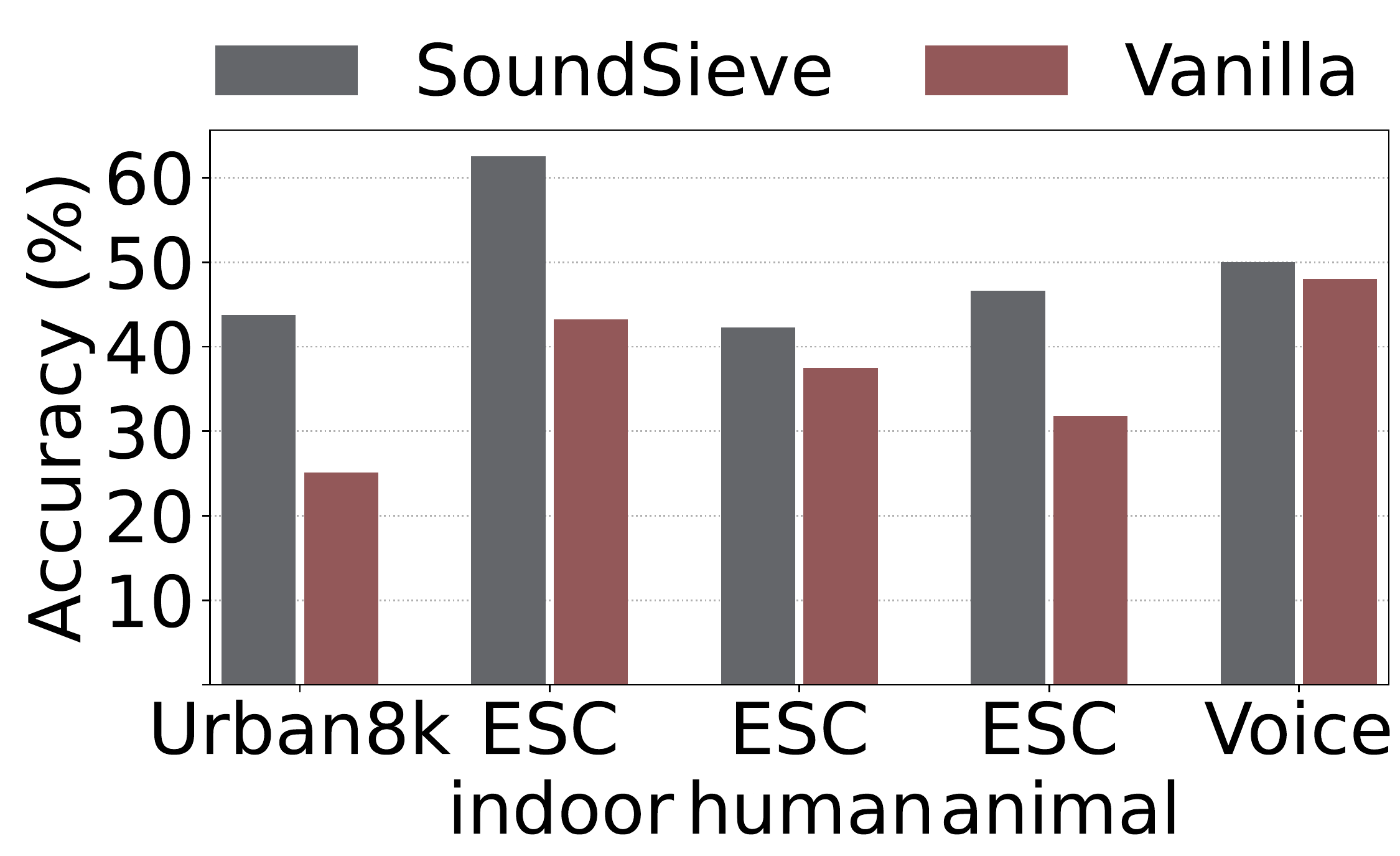}
    \caption{C=2}
  \end{subfigure}
  \hfill
  \begin{subfigure}[b]{0.225\textwidth}
    \includegraphics[width=\textwidth]{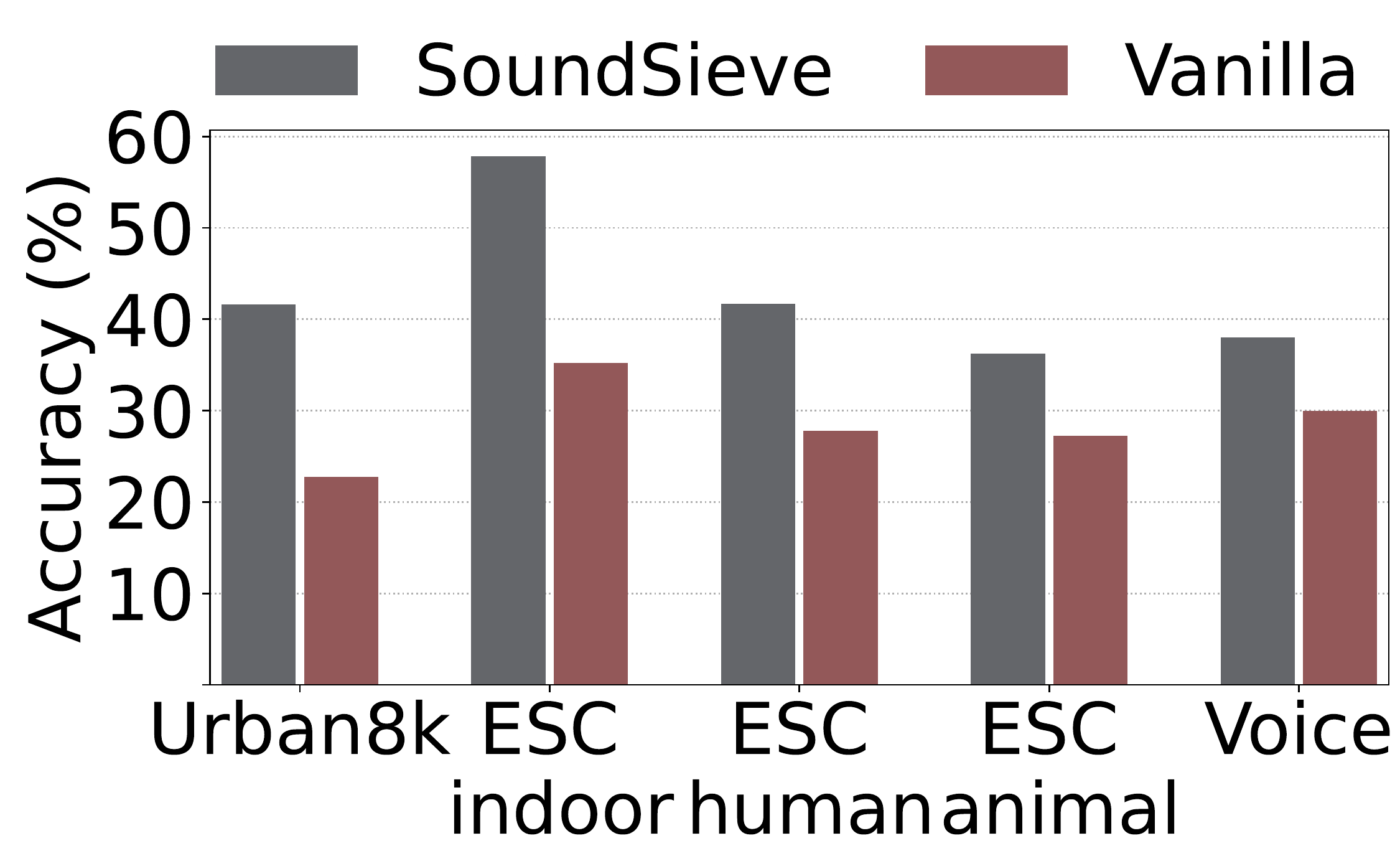}
    \caption{C=3}
  \end{subfigure}
  \caption{Comparison of SoundSieve and Vanilla approach at different energy harvesting rate}
  \label{fig:vanilla_vs_soundsieve}
\end{figure}

In Figure~\ref{fig:vanilla_vs_soundsieve}, we observe that \sys achieves up to 25\% higher classification accuracy than the uniform sampling strategy of the vanilla approach.

\subsection{Effect of Audio Length}
\begin{figure}[!htb]
  \centering
  \begin{subfigure}[b]{0.225\textwidth}
    \includegraphics[width=\textwidth]{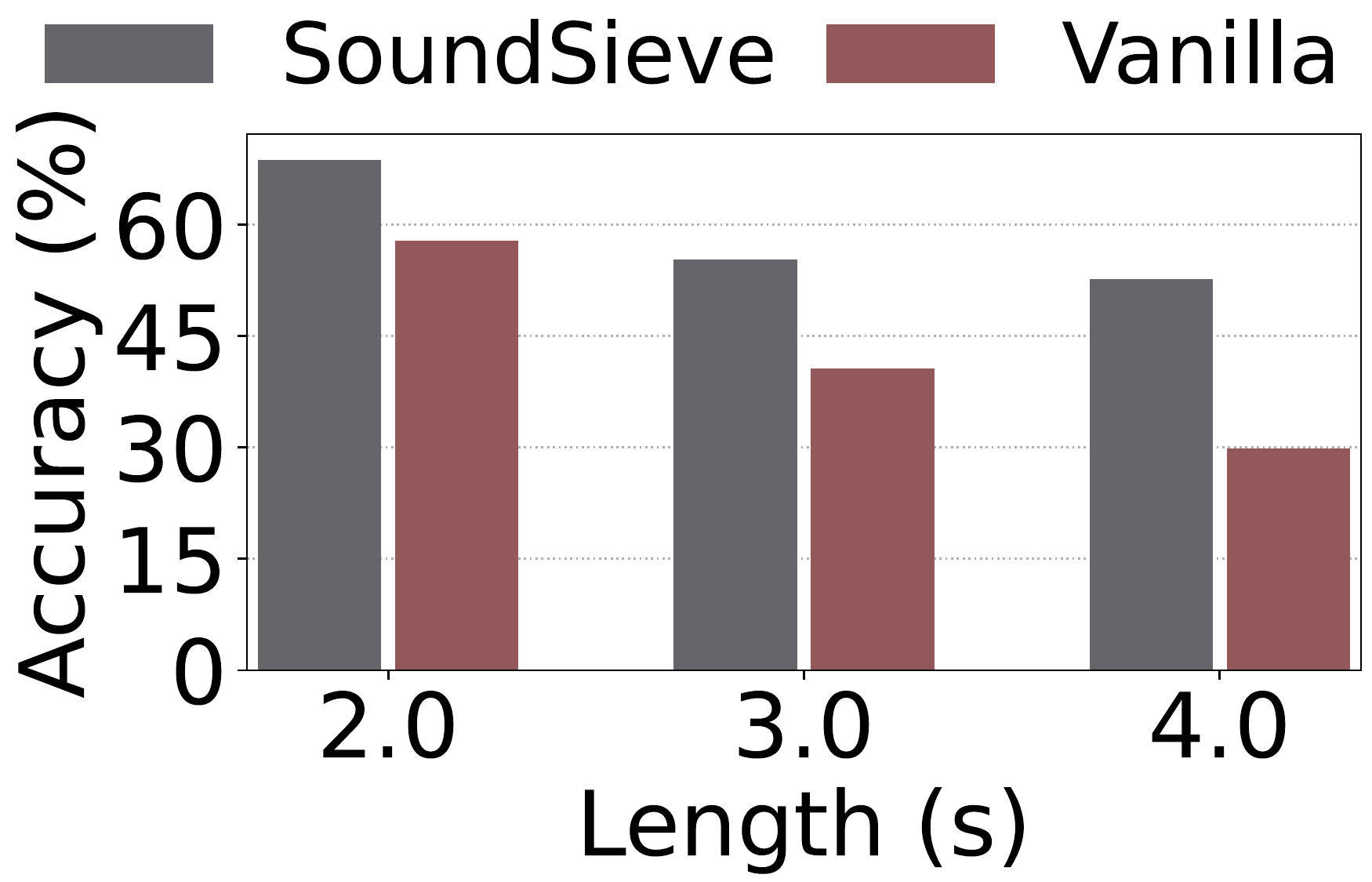}
    \caption{C=1}
  \end{subfigure}
  \hfill
  \begin{subfigure}[b]{0.225\textwidth}
    \includegraphics[width=\textwidth]{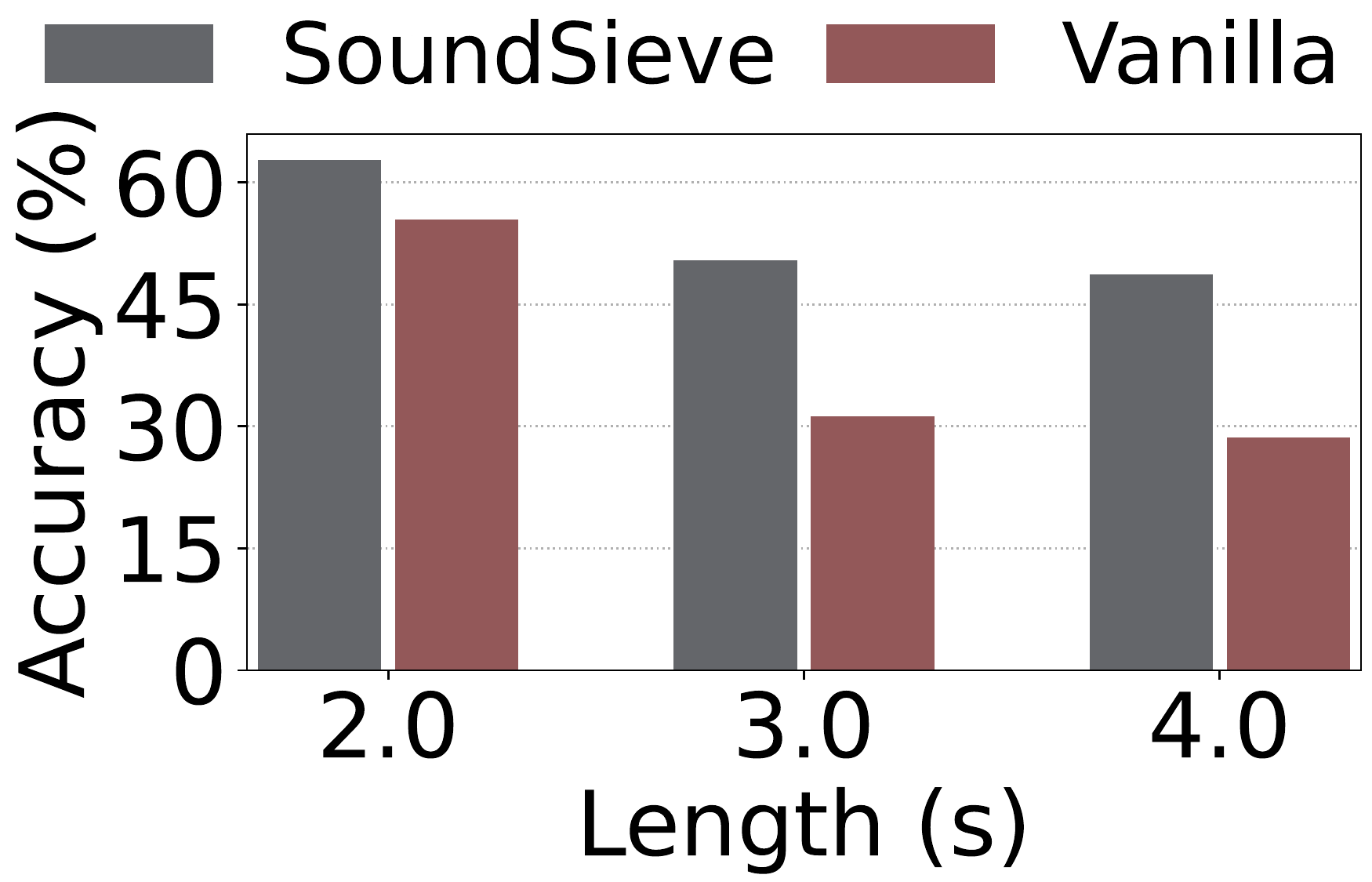}
    \caption{C=1.5}
  \end{subfigure}
  \\
  \begin{subfigure}[b]{0.225\textwidth}
    \includegraphics[width=\textwidth]{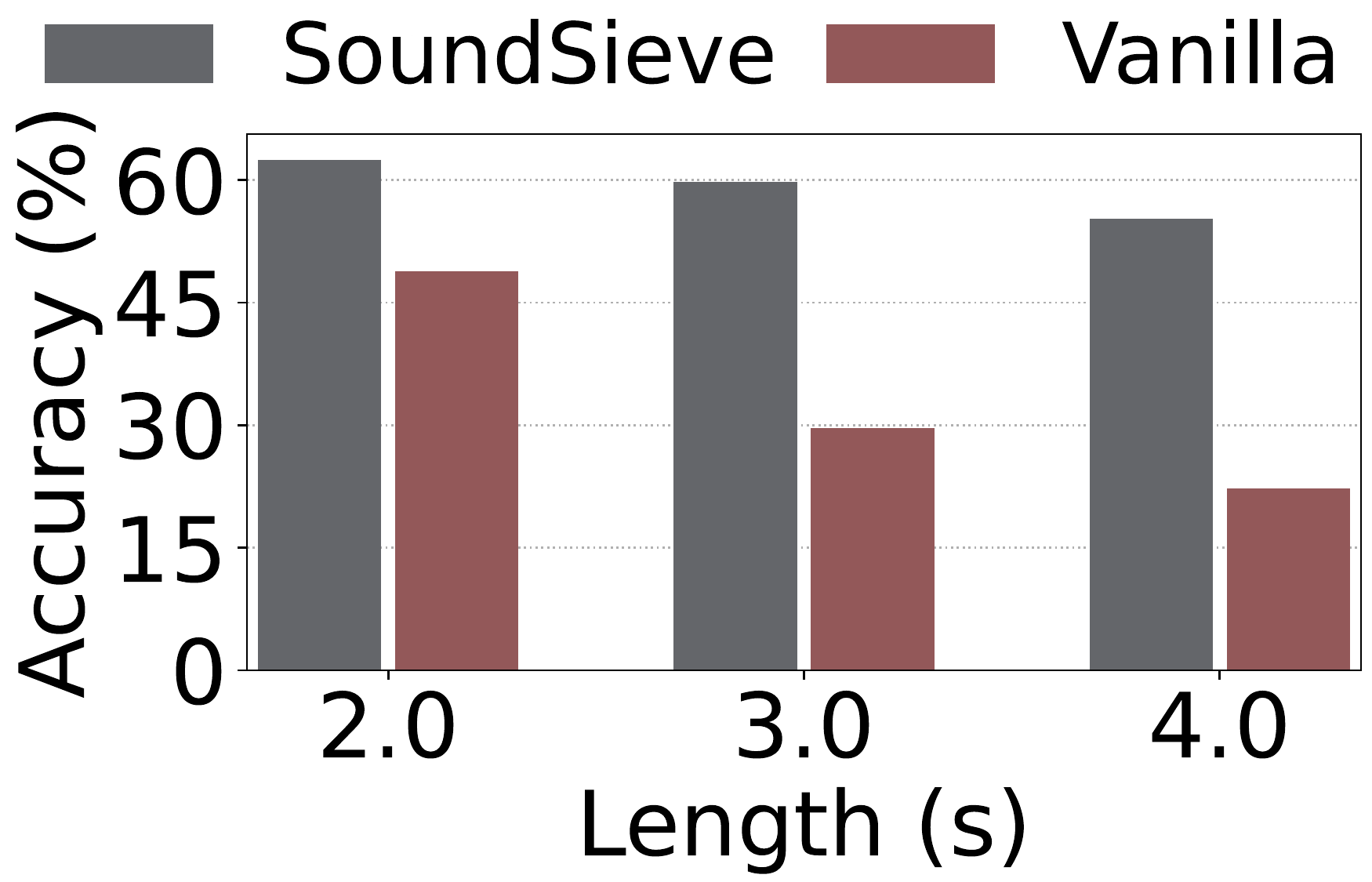}
    \caption{C=2}
  \end{subfigure}
  \hfill
  \begin{subfigure}[b]{0.225\textwidth}
    \includegraphics[width=\textwidth]{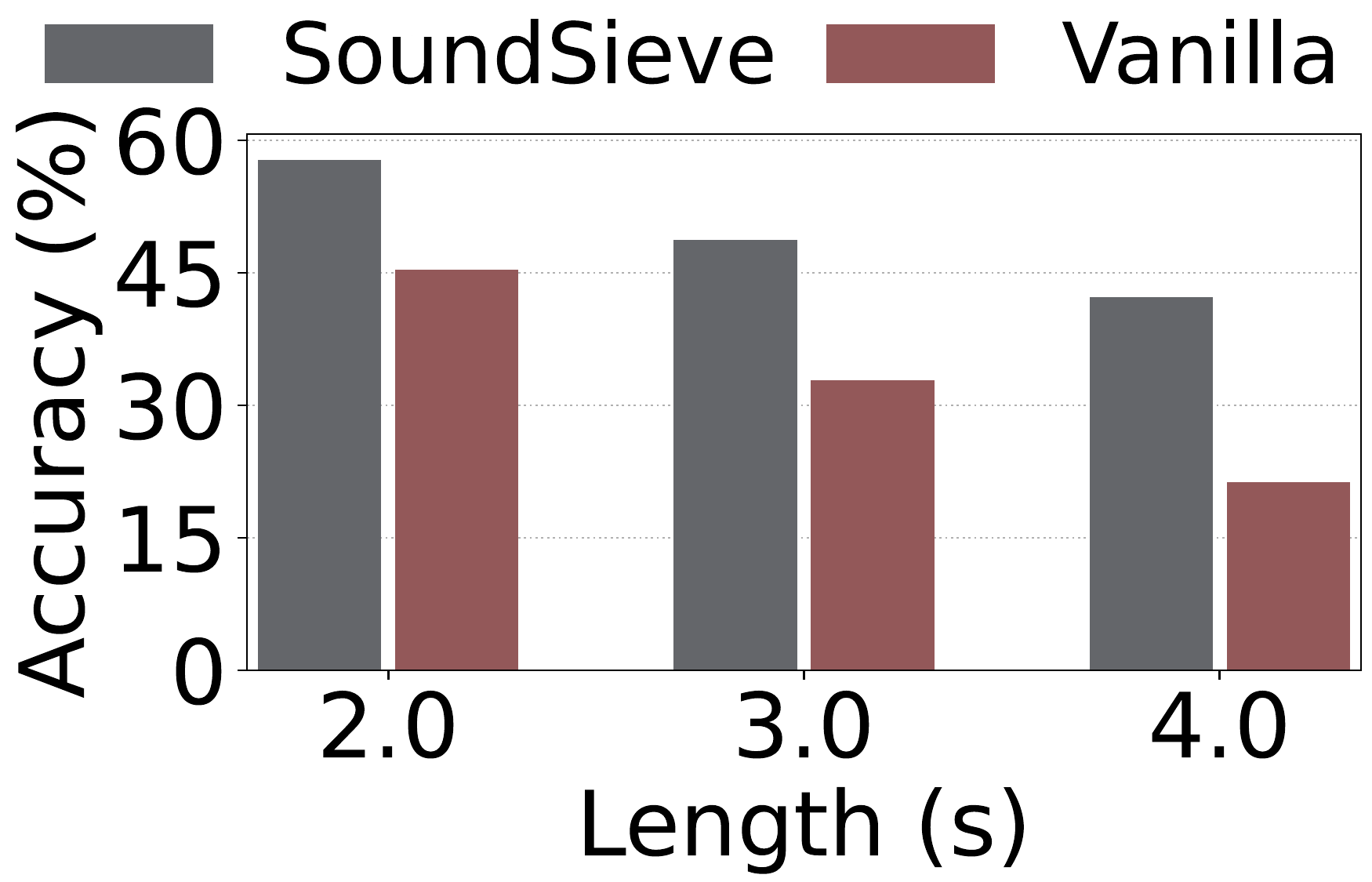}
    \caption{C=3}
  \end{subfigure}
  \caption{Effect of audio length at different energy harvesting rates.}
  \label{fig:length}
\end{figure}

\begin{figure}[!tbp]
  \centering
  \begin{subfigure}[b]{0.225\textwidth}
    \includegraphics[width=\textwidth]{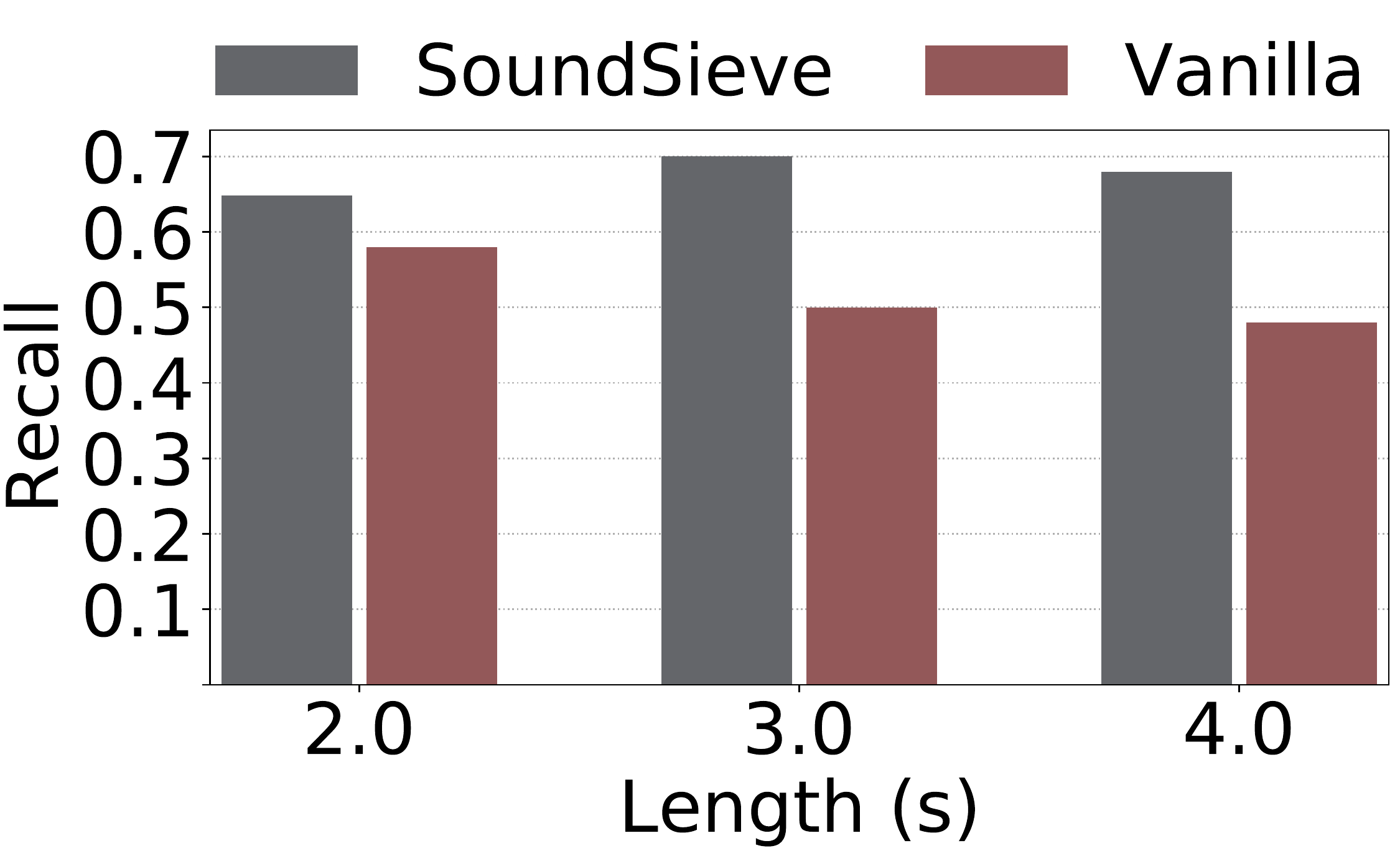}
    \caption{C=1}
  \end{subfigure}
  \hfill
  \begin{subfigure}[b]{0.225\textwidth}
    \includegraphics[width=\textwidth]{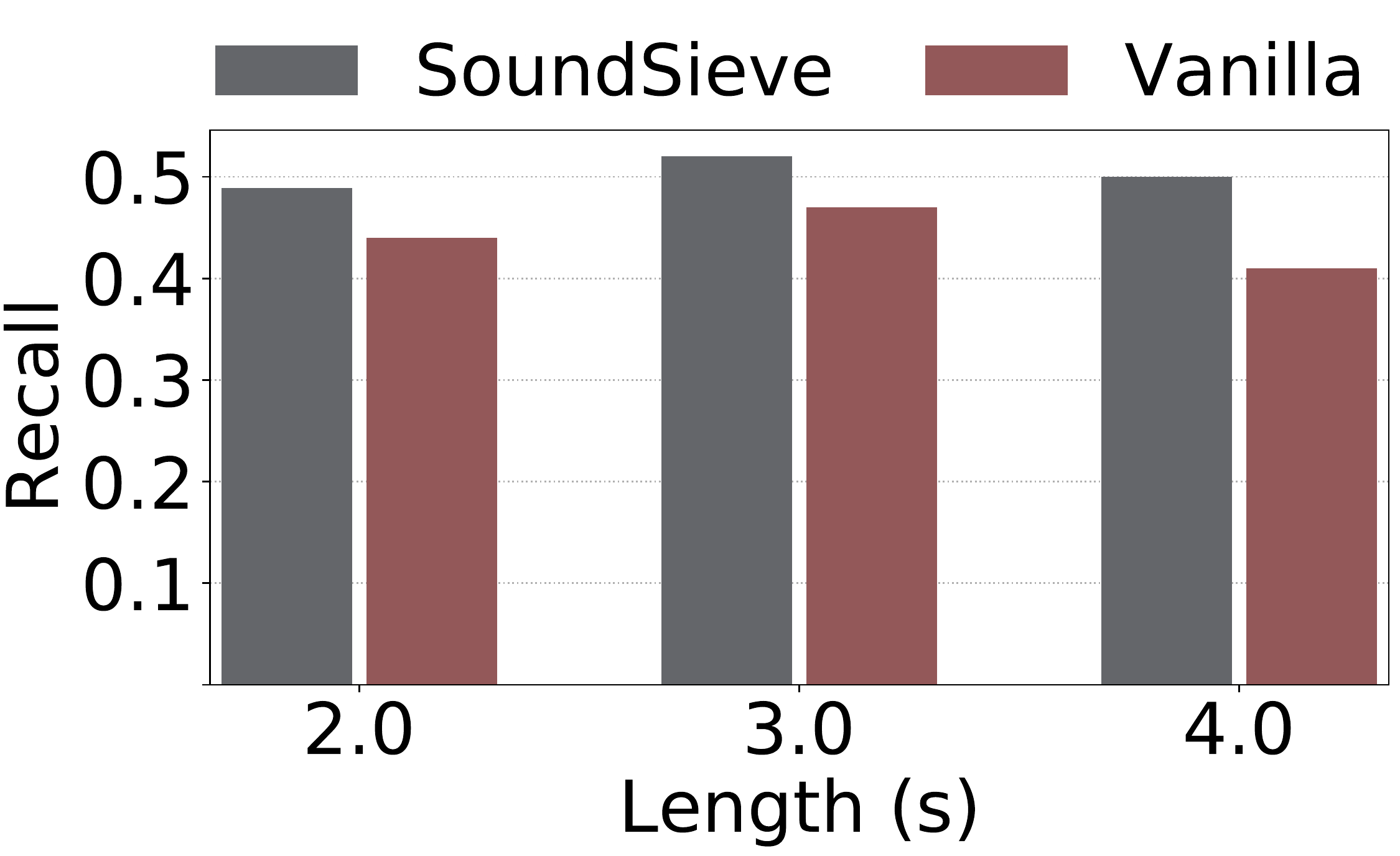}
    \caption{C=1.5}
  \end{subfigure}
  \\
  \begin{subfigure}[b]{0.225\textwidth}
    \includegraphics[width=\textwidth]{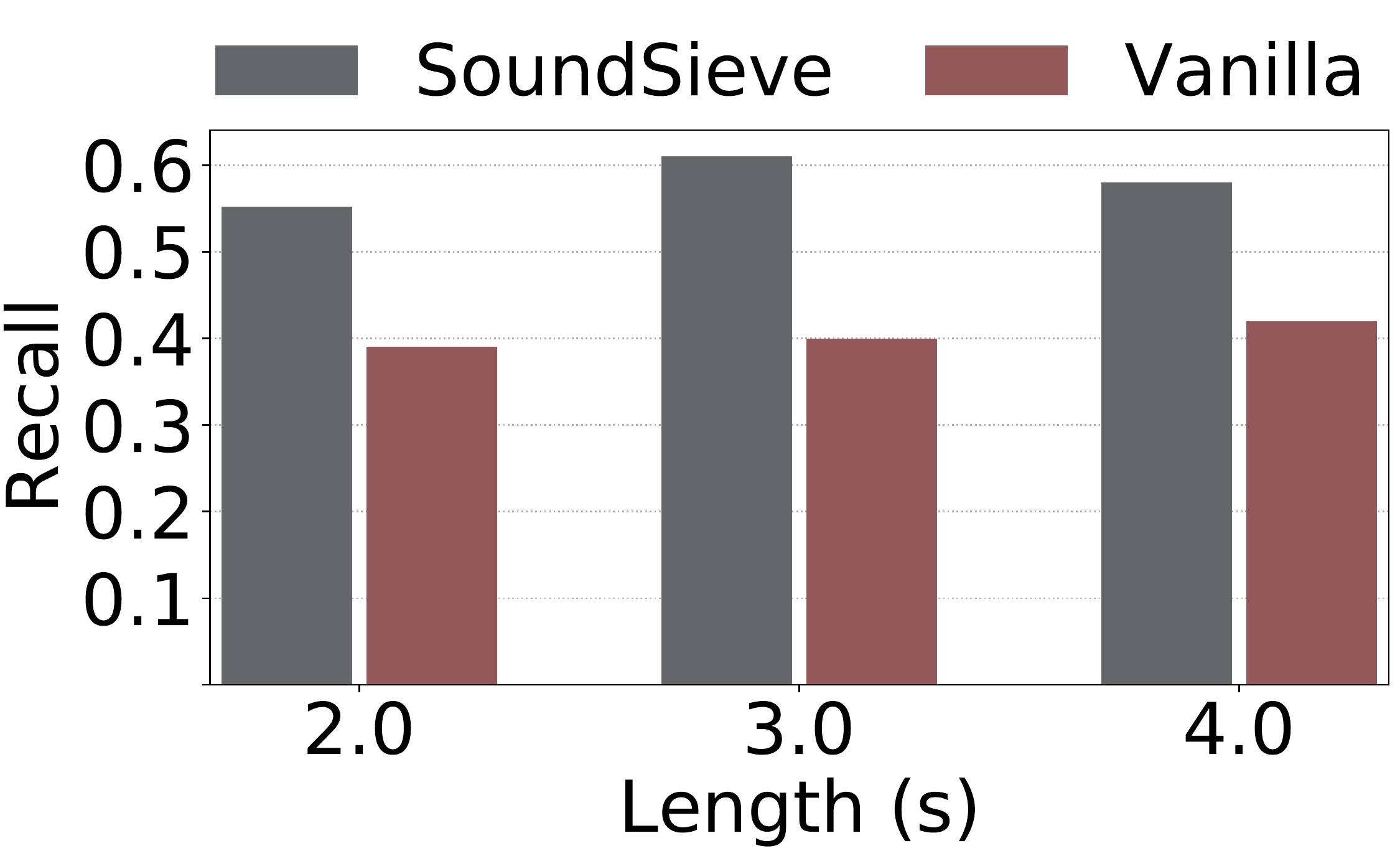}
    \caption{C=2}
  \end{subfigure}
  \hfill
  \begin{subfigure}[b]{0.225\textwidth}
    \includegraphics[width=\textwidth]{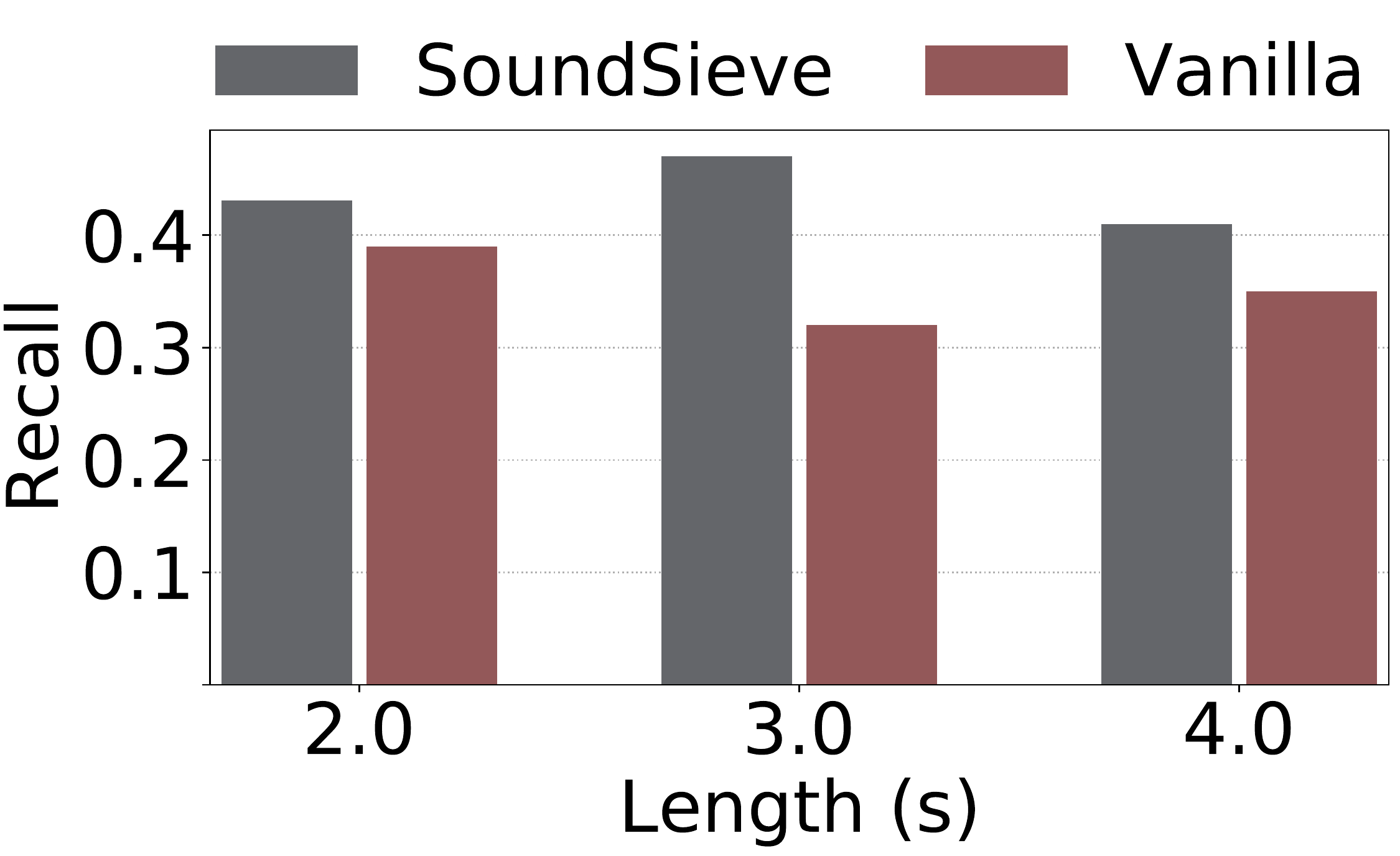}
    \caption{C=3}
  \end{subfigure}
  \caption{Comparison of recall of the most informative audio segments for different audio lengths and energy harvesting rates}
  \label{fig:recall2}
\end{figure}

We vary the length of audio clips from 2--4 seconds to test the robustness of \sys.
In Figure~\ref{fig:length}, we observe that \sys outperforms vanilla across all energy harvesting rates for all audio lengths. This is because as the length of the audio increases, the intermittent system goes through more charge and discharge cycles, and consequently misses more audio segments. Sensing more informative segments becomes crucial as the length of the audio increases. This is evident in Figure~\ref{fig:recall2} which compares their recall, i.e., the coverage of the most informative blocks by \sys and the vanilla. The recall for \sys remains consistent for audio clips of all lengths while it gets worse for the vanilla. 

Note that $80\%$ of the audio clips used in this experiment are about 2 seconds long and there are very few audio clips that are more than 3 seconds long. This creates an imbalance in the dataset and causes a drop in the accuracy as the audio length increases.

\subsection{Effect of Segment Size}
\label{sec:segsize}

\begin{figure}[!htb]
    \centering
    \includegraphics[width=\linewidth]{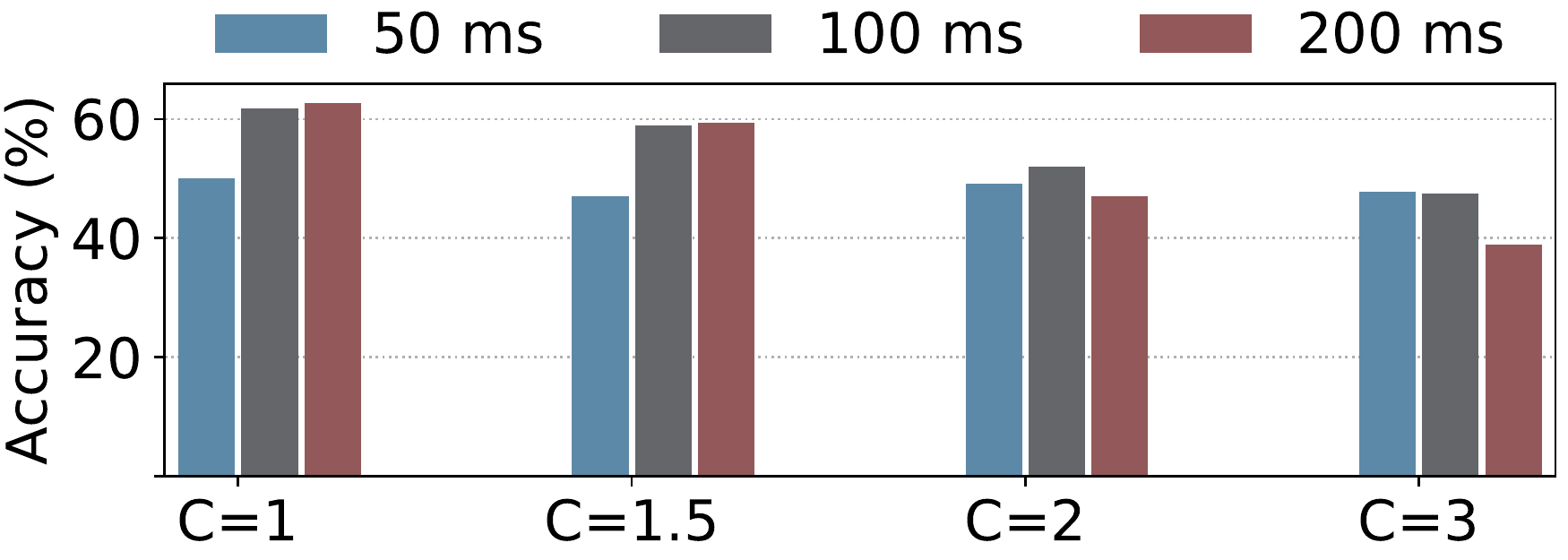}
    \caption{Performance of \sys at different segment size}
    \label{fig:segment}
\end{figure}

\sys analyzes and processes audio in the units of fixed-length segments. In this experiment, we study how the segment size affects \sys's classification accuracy under different energy harvesting conditions. We use the ESC-50~\cite{piczak2015esc} dataset for this experiment. The result is summarized in Figure~\ref{fig:segment}.

We observe that when the percentage of missing data is less, i.e., $C \le 1.5$, the classification accuracy increases as we increase the length of each segment. This is because, with larger segments, the system gets a longer window of audio signals to compute the Fourier transform and extract useful acoustic features. In contrast, when the percentage of missing data is very high, e.g., $C>2$, the system is highly likely to completely miss important segments if the segment length is large. Large segments give the system less control over its sampling decisions, which causes a drop in classification accuracy when a large portion of the data is missing. This is why we see that the trend is reversed for higher values of $C$ in Figure~\ref{fig:segment}. From this experiment, we learn that it is better to use shorter segments when the percentage of missing data is very high, and larger segments when the percentage of missing data is low in an intermittent system. 

\subsection{Beyond Audio}

Although \sys is specifically designed for audio signals, we conduct a dataset-driven, limited-scale study to understand its potential in other types of signals.

\begin{figure}[!htb]
    \centering
    \includegraphics[width=\linewidth]{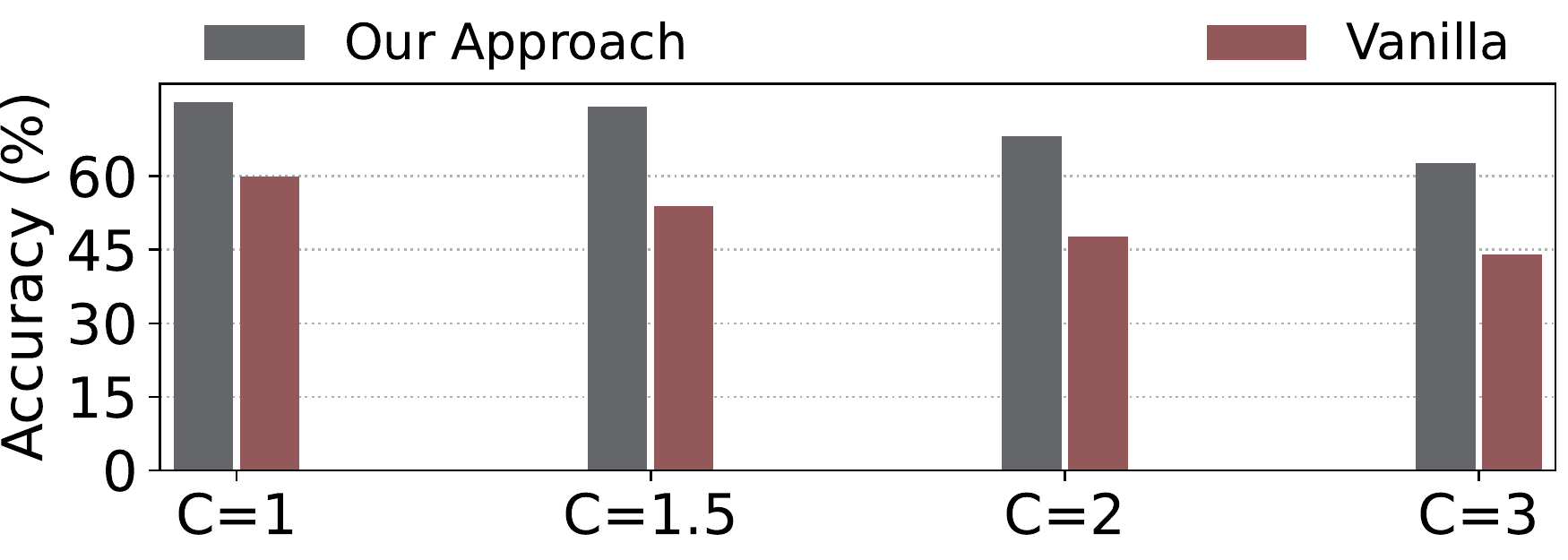}
    \caption{Comparison between our approach and the vanilla algorithm on UCI-HAR~\cite{ucihar} dataset.}
    \label{fig:imu}
\end{figure}

We use the UCI-HAR~\cite{ucihar} dataset, which contains accelerometer and gyroscope readings from 30 participants performing six types of activities. The dataset is split into train~($70\%$) and test~($30\%$) sets. Figure~\ref{fig:imu} compares the classification accuracy of our approach with that of the vanilla approach for different values of $C$. We observe that our approach achieves $12\%-20\%$ more accuracy than the Vanilla approach, and the gap increases when there are more missing data -- showing the resilience of the proposed technique in extreme energy harvesting conditions.

\section{System Implementation}

\subsection{Hardware Setup} 
We use TI MSP430FR5994~\cite{msp430} MCU having 256KB FRAM, 8KB SRAM, direct memory access (DMA), and an operating voltage range of 1.8V to 3.3V at 8MHz CPU clock speed. 

We use both solar and RF harvesters. The solar panel has polycrystaline solar cells~\cite{solar_panel} that outputs up to 5V at 40mA, which is regulated to 3.3V using~\cite{ltc3105}. The RF harvester consists of a TX91501 powercaster transmitter and a P2110 powerharvester receiver. A $470\mu F$ capacitor is used for energy storage.   

A PMM-3738-VM1010-R piezoelectric MEMS microphone is used for audio sensing. This microphone has a zero-power listening mode which allows it to continuously sense for audio event triggers at extremely low-power. Whenever an audio event is triggered, the microphone sends an interrupt to the MSP430 and the microphone is set to active listening mode. During this active mode, the system continuously samples audio at 5KHz, performs FFT, executes the first convolution layer of the classifier in the frequency axis and writes the data to the FRAM  using the low energy accelerator (LEA) and direct memory access (DMA).

\begin{figure}[!htb] 
    \centering
    \subfloat[Solar Harvester Setup]{
        \includegraphics[width = 0.233\textwidth]{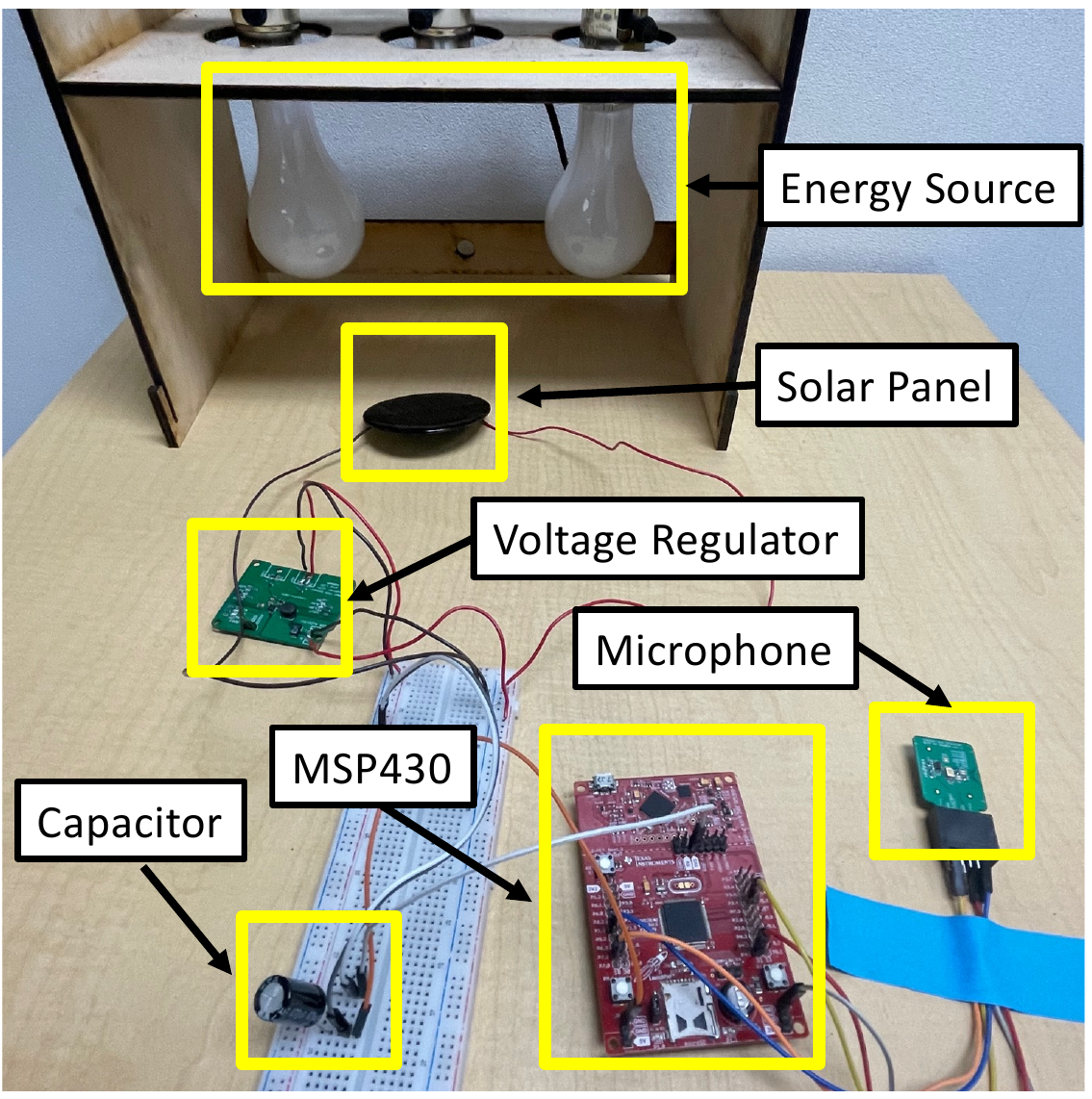} 
    }
    \subfloat[RF Harvester Setup]{
        \includegraphics[width = 0.233\textwidth]{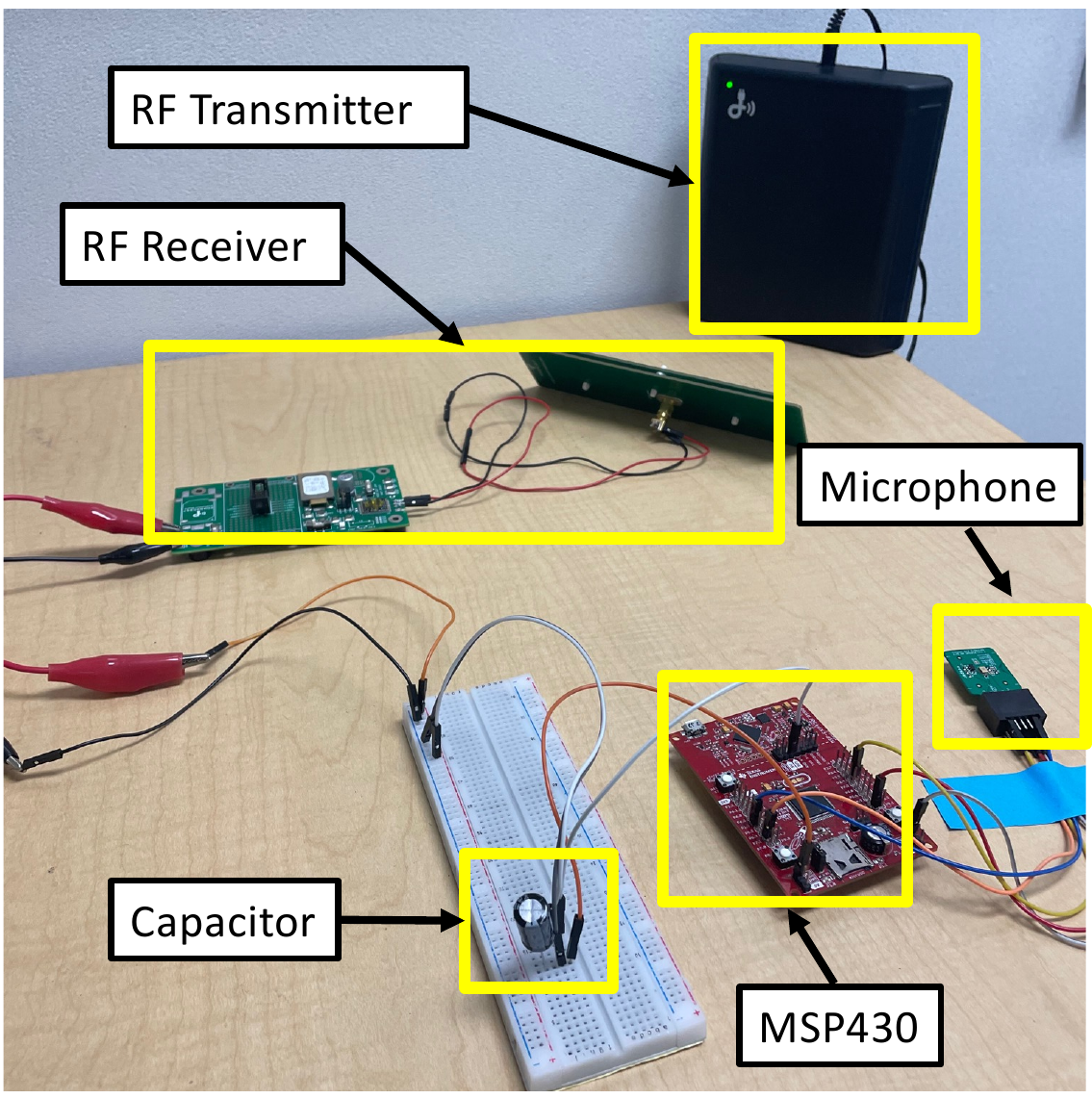} 
    }
    \caption{Hardware Setup. }
    \label{fig:hardware_real}
\end{figure} 

\subsection{Embedded System Software} 

We modify an open source Python tool ~\cite{Capuchin} to convert a TensorFlow model to a C header file which contains the weights and the neural network architecture. This header file is combined with platform-specific C implementation of neural network modules to obtain the audio classifier. We develop a complete C program that implements audio sensing, pre-processing, sampling, imputation, and classifier modules, and cross-compile to produce executable binary for the target system. We employ the static checkpoint calls similar to~\cite{mithreum}, where we profile the energy consumption at different states of sensing and when the capacitor voltage drops below 2.2V, the complete system state is copied to the FRAM.

\subsection{Microbenchmarking} 

\parlabel{Execution Tracing.} The intermittent system senses audio using a low-power microphone while it harvests energy from a RF harvester. We trace the operating voltage of the capacitor using a monitor~\cite{discovery}. After sensing each segment, \sys decides whether to sense the next segment or to skip some segments, and wait in low-power mode to conserve energy and to fill up the energy buffer. This decision is made based on the current energy level, global segment importance, and local segment importance -- predicted based on the current segment that is being observed. 

\begin{figure}[!htb]
    \centering
    \includegraphics[width=0.49\textwidth]{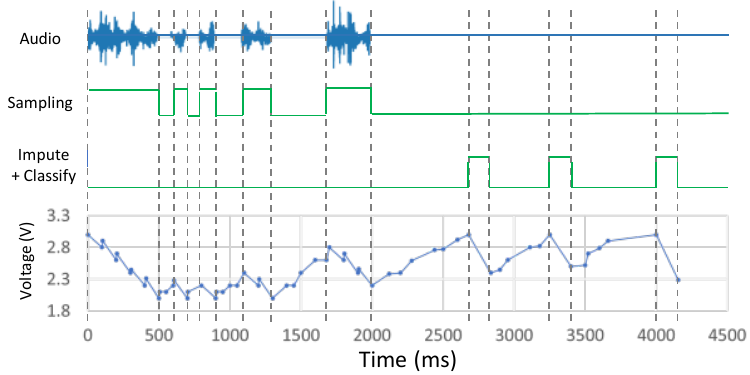}
    \caption{Execution steps of \sys}
    \label{fig:execution}
\end{figure}

A breakdown of all the steps of \sys for classifying one audio clip is visualized in Figure~\ref{fig:execution}. A breakdown of the sampling decisions is shown in Table~\ref{tbl:sampleexecution}. Initially, when there is no audio event to sense, the microphone remains in the low-power mode, and  the system periodically wakes up and checks its energy harvesting rate. Using this information, a sampling strategy based solely on the global importance scores and energy harvesting rate is created. This step corresponds to the first row of Table~\ref{tbl:sampleexecution}. When an audio event triggers, the system wakes up and starts sensing and processing the first audio segment. At the end of processing this segment, the system has the local importance scores of next 5 blocks as well as the global sampling strategy. If the local importance score of a segment is higher than a threshold and there is enough energy to spend, the system samples that segment; otherwise, it waits for a more informative segment according to the global sampling strategy. For example, at time $t_2$, even though the $x_5$ has the highest importance score, the system decides to sense $x_3$ since its importance score was greater than the pre-determined threshold. After the audio event ends, the system starts executing the DNN to classify the audio event.

\begin{table}[!htb]
 \resizebox{\linewidth}{!}{
\begin{tabular}{|l|l|l|l|}
\hline
 \textbf{Time} & \textbf{State} & \textbf{Segments by Importance} & \textbf{Decision}\\
 \hline
 $t_0$ & init & $(x_2, x_7, x_5, x_8, x_1, \cdots )$ & Create Initial Sampling Strategy \\   
 $t_1$ & sample & $(x_2, x_5, x_3, x_6, x_4)$ & Sense at $t_2$ \\
 $t_2$ & sample & $\{x_5, x_3, x_7, x_4, x_6\}$ & Sense at $t_3$ \\
 $t_3$ & sample & $\{x_7, x_4, x_8, x_5, x_6\}$ & Sense at $t_4$ \\
 $t_4$ & sample & $\{x_5, x_7, x_8, x_9, x_6\}$ & Sense at $t_5$ \\
 $t_5$ & sample & $\{x_8, x_7, x_9, x_{10}, x_6\}$ & Wait until $t_7$ \\ 
 $t_6$ & wait &  & Wait until $t_7$ \\
 $t_7$ & sample & $\{x_9, x_8, x_{11}, x_{12}, x_{10}\}$ & Wait until $t_9$ \\
 $t_8$ & wait &  & Wait until $t_9$ \\
 $t_9$ & sample & $\{x_{13}, x_{12}, x_{11}, x_{10}, x_{14}\}$ & Wait until $t_{12}$ \\
 $t_{10}$ & wait &  & Wait until $t_{12}$ \\
 $t_{11}$ & wait &  & Wait until $t_{12}$ \\
 $t_{12}$ & sample & $\{x_{13}, x_{16}, x_{14}, x_{17}, x_{15}\}$ & Wait until $t_{13}$ \\
 $t_{13}$ & sample & $\{x_{14}, x_{18}, x_{15}, x_{16}, x_{17}\}$ & Wait until $t_{18}$ \\
 \hline
\end{tabular}}
 \caption{Sampling strategy for an example audio clip}
\label{tbl:sampleexecution}
\end{table}

\parlabel{Overhead Measurement.} Table~\ref{tbl:time} summarizes the time and energy overhead breakdown of on-device audio processing. Sensing and sampling decisions are made sporadically throughout the duration of the audio. The time overhead of the sampling algorithm is negligible since FFT and feature extraction for sampling decision are computed in parallel. The DNN runs when the microphone signals the end of audio. It takes less than a second (uninterrupted) to classify the audio---which can be interrupted 2-3 times due to intermittence.        

\begin{table}[!htb]
 \resizebox{0.48\textwidth}{!}{
\begin{tabular}{|l|l|l|}
\hline
 \textbf{Module} & \textbf{Time (ms)} & \textbf{Energy ($\mu$J)}\\
 \hline
 Sensing + Preprocess & 100 & 510 \\
 Sampling Decision &  7 & 60\\
 DNN CONV (3 layers) & 707 & 5850 \\
 DNN FC layer & 5 & 100 \\
 \hline
\end{tabular}}
 \caption{Execution time and energy overhead breakdown}
\label{tbl:time}
\end{table}

\begin{figure*}[!thb]
    \centering
    \includegraphics[width=\textwidth]{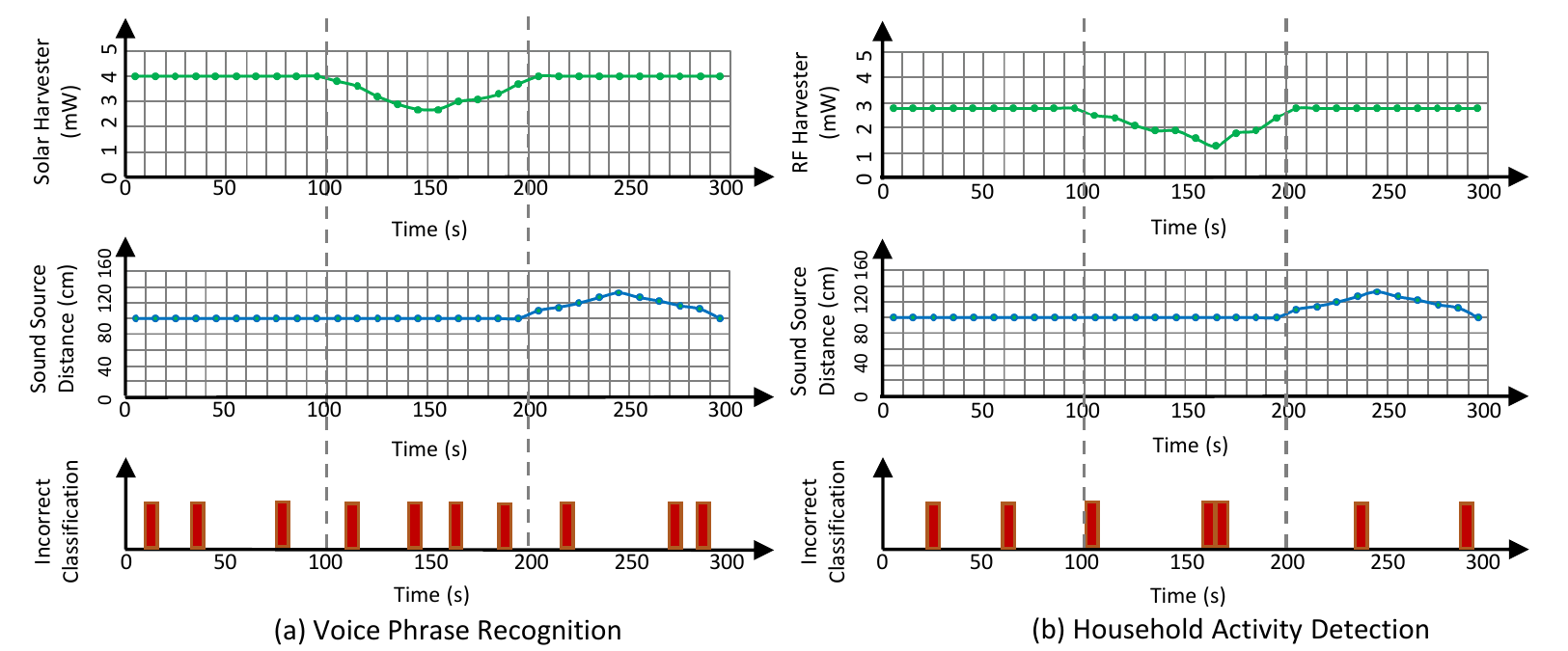}
    \caption{Effect of mobility and energy harvester in a real intermittent system.} 
    \label{fig:deployment30}
\end{figure*}

\section{Evaluation on Real System}

\subsection{Experimental Setup}

\parlabel{Applications.} After validating the sampling technique, imputation method, and the classifier using dataset-driven experiments, we design two real-world deployment scenarios using both solar and radio frequency (RF) harvesters. We implement two realistic applications: household activity detection, and voice phrase recognition. Each application has 10 classes of audio events and their lengths are shown in table~\ref{tab:my_label}. We use the DNN architecture shown in the Table~\ref{tbl:netarch} for classification. The DNN is split into three sub-networks to enable intermittent execution. After executing each sub-network, the intermediate results are stored in the FRAM. Then the system waits in the low-power mode while the capacitor charges, and periodically monitors the energy level. Finally, when enough energy is accumulated, the next sub-network of the DNN begins execution.   

\begin{table}[!thb]
    \centering
    \begin{tabular}{|c|c|c|}
        \hline
        Application & Classes & Length (s)\\
        \hline
        Voice Phrases & 10 & 1-3\\
        Household Activities& 10 & 2-4\\
        \hline
    \end{tabular}
    \caption{Two applications used for real system deployment. Voice phrases include ten different phrases such as ``go left", ``go right" etc. Household activities include sound events such as brushing teeth, pouring water, washing machine etc.}
    \label{tab:my_label}
\end{table}

\parlabel{Methodology.} For each application, we randomly select 30 clips for test -- 1 clip for each of the 10 sound categories. We create a playlist with these 30 clips and play each clip from a mobile device every 10 second. Hence, the duration of the experiment is 300 seconds. Playing the sounds from a mobile device makes it possible to repeat the experiments and compare the results under different operation conditions such as mobility and energy variation.       

To account for mobility and variation in harvested energy, we divide the 300 seconds into three 100 seconds phases. In the first phase, we keep both the sound source and the energy source fixed relative to the microphone. In the second and the third phases, we move the sound source and the energy source, respectively, at constant velocity. We use the solar setup for the voice application and the RF setup for the household application.

\subsection{Experimental Results} 
\parlabel{Effect of Harvested Energy.}
In order to observe the effect of harvested energy we increase the distance between the RF harvester and the RF source when harvesting energy from RF, and change the light intensity when harvesting energy from solar. In Figure~\ref{fig:deployment30} between 100 seconds to 200 seconds, we observe that the classifier is able to maintain similar classification performance across both application for varying energy harvesting rates. During this period, we observe 4 out of 10 incorrect classifications for voice phrases and 3 out of 10 incorrect classifications for household activities, resulting in an accuracy of 60\% and 67\%, respectively.  

\parlabel{Effect of Sound Source Mobility.} We vary the distance between the sound source and the low-power microphone from $40$ cm to $60$ cm. The microphone is able to consistently wake up to the sound event and the classification accuracy remains similar as shown in Figure~\ref{fig:deployment30} from 200 seconds to 300 seconds.
During this period, we observe 3 out of 10 incorrect classifications for voice phrases and 2 out of 10 incorrect classifications for household activities, resulting in an accuracy of 67\% and 80\%, respectively.  

\parlabel{Effect of Audio Length.} We experiment with audio clips of varying lengths -- ranging from 1 second to 3 seconds for voice phrases, and 2 seconds to 4 seconds for household activity sounds. Since \sys is able to identify more informative segments in an audio, the classification performance remains consistently similar for audio clips of all lengths. Overall, we observe 10 out of 30 incorrect classifications for voice phrases and 7 out of 30 incorrect classifications for household activities, resulting in an accuracy of 67\% and 76\%, respectively.
\section{Related Work}
\parlabel{Intermittent Computing.} Intermittently powered systems experience frequent power failure that resets the software execution and results in repeated execution of the same code, and inconsistency in non-volatile memory. Previous works address the progress and memory consistency using software check-pointing~\cite{ransford2012mementos, maeng2018adaptive, hicks2017clank, lucia2015simpler, van2016intermittent, van2016intermittent, jayakumar2014quickrecall, mirhoseini2013automated, bhatti2016efficient}, hardware interruption~\cite{balsamo2015hibernus, balsamo2016hibernus++, mirhoseini2013idetic}, atomic task-based model~\cite{maeng2017alpaca, colin2016chain, colin2018termination}, non-volatile processors (NVP)~\cite{ma2017incidental, ma2015architecture}, and adaptive inference~\cite{wu2020intermittent, li2021developing, li2021implementation, li2022energy}. 

Recently SONIC~\cite{gobieskiintermittent, gobieski2019intelligence, lee2019intermittent} proposes a unique software system for intermittent execution of deep neural inference combining atomic task-based model with loop continuation. None of the previous works perform intelligent sampling like \sys.

\parlabel{Modeling Energy Harvesting Systems.}
~\cite{san2018eh} analytically model the trade-off associated with backing up data to maximize forward propagation. Even though energy harvesting system for a specific energy source has been analyzed and modeled before~\cite{crovetto2014modeling, sharma2018modeling, jia2018modeling}, none of the prior works focus on modeling the intermittence of sensor data on an intermittent system.  

\parlabel{Intermittent Audio Sensing.}
In~\cite{continuous} authors demonstrated audio sensing in energy harvested devices, however, they only sense audio events of 283ms in length, which is not adequate for many real-world applications. In~\cite{9747219}, authors simulated intermittency in audio sensing, however, they run post-processing and speech recognition networks on high-end devices without considering any computation and energy constraints. 

\parlabel{Adaptive Informative Sampling.}
Resource constraint systems often need to monitor and sense large volume of data including video~\cite{bappy2019exploiting, fundus}, RF~\cite{munir2022carfi, fung2019coordinating} and audio~\cite{kemna2016adaptive}. In order to perform time and cost effective monitoring of the environment, many tiny robots and automated machines utilize adaptive informative sampling~\cite{kemna2018board, kemna2016adaptive, fung2019coordinating}. These robots model the environmental phenomena using Gaussian process regression~(GPR)~\cite{wilson2011gaussian}. However GPR is an useful technique for modeling spatial uncertainty, not temporal uncertainty. 
\section{Conclusion}
In this paper, we develop an intermittent audio sensing system that classifies audio events that are longer than the discharging cycle of the system. We leverage inherent properties of audio signals to design a lightweight audio sampler, an imputation method, and a classifier network to classify audio events. We show the effectiveness of the system using both dataset-driven and real-world evaluations. In the future, we plan to explore the possibility of designing content-aware intermittent sensing and classification systems that deal with complex data such as camera images.

\begin{acks}
This work was supported, in part, by grants NIH 1R01LM013329-01 and NSF 2047461.
\end{acks}

\balance
\bibliographystyle{ACM-Reference-Format}
\bibliography{references}

\end{document}